# A computational paradigm for dynamic logic-gates in neuronal activity

Amir Goldental[1†], Shoshana Guberman[1,2†], Roni Vardi[2†] and Ido Kanter[1,2]*

[1]*Department of Physics, Bar-Ilan University, Ramat-Gan 52900, Israel*
[2]*Gonda Interdisciplinary Brain Research Center and the Goodman Faculty of Life Sciences, Bar-Ilan University, Ramat-Gan 52900, Israel*

[†] These authors equally contributed to this work

**\* Correspondence:** *Ido Kanter, Gonda Interdisciplinary Brain Research Center, Bar-Ilan University, Ramat-Gan 52900, Israel. E-mail: ido.kanter@gmail.com.*

## Abstract

In 1943 McCulloch and Pitts suggested that the brain is composed of reliable logic-gates similar to the logic at the core of today's computers. This framework had a limited impact on neuroscience, since neurons exhibit far richer dynamics. Here we propose a new experimentally corroborated paradigm in which the truth tables of the brain's logic-gates are time dependent, i.e. dynamic logic-gates (DLGs). The truth tables of the DLGs depend on the history of their activity and the stimulation frequencies of their input neurons. Our experimental results are based on a procedure where conditioned stimulations were enforced on circuits of neurons embedded within a large-scale network of cortical cells in-vitro. We demonstrate that the underlying biological mechanism is the unavoidable increase of neuronal response latencies to ongoing stimulations, which imposes a non-uniform gradual stretching of network delays. The limited experimental results are confirmed and extended by simulations and theoretical arguments based on identical neurons with a fixed increase of the neuronal response latency per evoked spike. We anticipate our results to lead to better understanding of the suitability of this computational paradigm to account for the brain's functionalities and will require the development of new systematic mathematical methods beyond the methods developed for traditional Boolean algebra.

## 1. Introduction

This year we are celebrating the 70[th] anniversary of the publication of the seminal work by Warren S. McCulloch, a neuroscientist, and Walter Pitts, a logician, entitled "A logical calculus of the ideas immanent in nervous activity" (McCulloch and Pitts, 1943). They attempted to understand how the brain could produce highly complex patterns by using many interconnected building blocks of the brain, the neurons. In their model, the brain is composed of Boolean entities functioning as threshold units. Such simplified units constitute pure and reliable logic-gates (e.g. AND, XOR), similar to the logic at the core of today's computers. The generalization of this simplified Boolean framework to include unreliable elements has emerged in 1956 by the innovative work of John von Neumann (Von Neumann, 1956). These concepts as well as the earlier pioneering work of Claude Shannon to simplify Boolean circuits (Shannon, 1938) are at the cornerstone of today's computational paradigm

(Turing, 1938).

The computational framework of McCulloch and Pitts had a tremendous impact on the development of artificial neural networks (Hopfield, 1982; Krogh, 2008; Qian, 2011; Gerstner et al., 2012; Gilja et al., 2012) and machine learning theory (Sutton and Barto, 1998; Hunt et al., 2012). Their concept triggered the next major development in theoretical neural networks when in 1958 Frank Rosenblatt introduced the concept of the perceptron (Rosenblatt, 1958), the prototypical linear classifier, which ever since has been theoretically investigated and generalized to more structured multi-layer and recurrent architectures (Litwin-Kumar and Doiron, 2012; Stoianov and Zorzi, 2012). Nevertheless, it is fair to conclude that the concept of simplified Boolean neurons had a limited impact on neuroscience, which exhibit much richer temporal dynamics (Izhikevich, 2006; Izhikevich and Hoppensteadt, 2009; Gal et al., 2010; Vardi et al., 2012a). Moreover, it appears that the brain is the most ineffective environment to implement such a Boolean logical operating system, comprised of *static* logic-gates (SLGs).

Seven decades after the proposed neuronal paradigm by McCulloch and Pitts, the fundamental concept of the computational abilities of the nervous system remains unclear (Hodges, 2012). On the one hand, one might conclude that the search for a comprehensive computational logic framework is irrelevant, as specialization in specific behavioral and perceptual tasks requires different "operating systems". On the other hand, it is evident that the "hardware" implementations of all complex brain tasks are composed of similar basic interconnected building blocks (neurons) having many features in common, which are enhanced and possibly dominant when operating as an ensemble (Abeles, 1991).

In the present study, we extend the recently demonstrated new experimentally corroborated paradigm in which the logical operations of the brain differ from the logic of computers (Vardi et al., 2013b). Unlike a burned logic-gate on a designed chip that consistently follows the same truth-table, here the functionality of the brain's logic-gates depend on the history of their activity, the stimulation frequencies of their input neurons, as well as on the activity of their interconnections. Our results are based on an experimental procedure where conditioned stimulations were enforced on circuits of neurons embedded within a large-scale network of cortical cells in-vitro (Marom and Shahaf, 2002; Morin, 2005; Wagenaar et al., 2006; Vardi et al., 2012b). We demonstrate that the underlying biological mechanism is the unavoidable increase of neuronal response latencies to ongoing stimulations (Aston-Jones et al., 1980; De Col et al., 2008; Ballo and Bucher, 2009; Gal et al., 2010; Soudry and Meir, 2012), which imposes a non-uniform gradual stretching of delays associated with the neuronal circuit (Kanter et al., 2011; Vardi et al., 2012a; Vardi et al., 2013a; Vardi et al., 2013c). To further support and expand the limited experimental results, we present a straightforward theoretical model based on the assumption of identical neurons with a constant increase in their neuronal response latency per evoked spike. This model, corroborated with simulations, allows us to explore the behavior of more complex structured neuronal DLGs in addition to SLG (Vogels and Abbott, 2005). We anticipate our results to be a starting point for larger scale in-vitro experiments and structured recurrent neuronal circuits, which will lead to a better understanding of the suitability of this computational paradigm to account for the brain's functionalities. In addition, this paradigm will require the development of new systematic methods and practical tools beyond the methods developed for traditional Boolean algebra (Chavesa, 2005; Nahin, 2012).

## 2.     Elastic response latency

## 2.1. Single neuron

The neuronal response latency, measured as the time-lag between a stimulation and its corresponding evoked spike, is one of the most significant time-dependent features at the single neuron level, and typically it is on the order of several milliseconds (Eccles et al., 1966; van Pelt et al., 2004; Ballo and Bucher, 2009; Gal et al., 2010; Vardi et al., 2012a). When stimulated repeatedly, a neuron exhibits a tendency to gradually stretch its stimulus-response delay over few milliseconds (Spira, 1976; Grossman et al., 1979; Thomson and West, 1993; Tal, 2001; Fuhrmann et al., 2002; Bakkum, 2008; Scroggs, 2008).

To exemplify this neuronal feature, stimulations at a rate of 10 Hz (**Figure 1A** (Vardi et al., 2013b)) were given to cultured cortical neurons that were functionally isolated from their network by pharmacological blockers of both excitatory and inhibitory synapses (see Appendix). The stimulated neuron responded with a very high reliability, resulting in a typical increase of a few milliseconds in the response latency over a few hundreds of repeated stimulations (**Figure 1A** (Vardi et al., 2013b)). Results indicate that the neuronal response latency increases by a few μs per evoked spike, which represents a finer time scale of cortical dynamics, μs, as discussed at (Vardi et al., 2012a). Specifically, one might notice three main trends of the response latency increase. For the first several stimulations there is a large increase in the neuronal response latency, in the order of several dozen μs per evoked spike (**Figure 1A** (Vardi et al., 2013b)). This state is followed by a fast decay to the second state, where the average increase in the neuronal response latency per evoked spike is only several μs, and the stretching of the neuronal response latency is roughly linear. The second state is the main contributor to the latency increase and lasts for a relatively long section of the stimulation period. In the presented experiment the second state starts after ~100 stimulations and lasts for approximately 550 stimulations, periods which vary across different neurons. Finally the neuron enters the third state, known as the intermittency phase (Gal et al., 2010; Vardi et al., 2012a), characterized by fluctuations around an average latency (starts after ~650 stimulation in the presented experiment). An apparent increase in the neuronal response latency to periodic stimulations can be observed for stimulation rates higher than ~3 Hz. Typically, the higher the stimulation rate, the larger the average increase of the response latency per evoked spike (Gal et al., 2010; Vardi et al., 2012a). This process is a fully reversible phenomenon and after a waiting time of a few seconds without stimulations, the response latency substantially decays and in a timescale of several minutes the initial response latency is completely restored.

The approximately linear increase in the neuronal response latency per evoked spike before entering the intermittent stage is at the center of our study. Consequently, the proposed theoretical methods are based on the approximation that the neuronal response latency increases by a constant value ($\Delta$) per evoked spike (identical for all neurons and time-independent).

## 2.2. Circuit level

To analyze the impact of dynamic neuronal response latency at a circuit level, we artificially generated conditioned stimulations over a circuit of neurons embedded within a large scale network of cortical cells in-vitro (see Appendix). Our first experimental design consisted of a chain of two neurons (**Figure 1B** (Vardi et al., 2013b)). Neuron A is stimulated at a rate of 10 Hz and the initial time-gap between consecutive evoked spikes of neurons A and B is set to $\tau_{AB}$=80 ms (neuron B is stimulated 80-$L_B(0)$ ms after an evoked spike of neuron A, where $L_B(0)$ stands for the initial response latency of neuron B) (**Figure 1B** (Vardi et al., 2013b)). After ~ 270 stimulations the response latency of neuron B increases by ~2 ms, thus resulting in an increase of the delay, $\tau_{AB}\approx 82$ ms.

The increase in the delays of the neuronal chain has an accumulative effect, as a result of the increase in the neuronal response latencies of the neurons comprising the chain (**Figure 1C** (Vardi et al., 2013b)). More neurons in a chain lead to a faster and greater increase of the entire delay of the chain. In order to compare results of two-neuron and five-neuron chains, a chain of 5 neurons (A,B,C,D,E) was examined. $\tau_{AE}$ was set to 80 ms, resulting in an initial time-gap of 80 ms between evoked spikes of neurons A and E, where $\tau_{AE}=\tau_{AB}+\tau_{BC}+\tau_{CD}+\tau_{DE}$. In the presented experiment the initial delays between consecutive neurons were selected to be equal, however, results are robust to arbitrary delays summing up to $\tau_{AE}$. After ~270 stimulations of neuron A, where each stimulation results in an evoked spike of neuron E, the stretching of $\tau_{AE}$ is about 6 ms (**Figure 1C** (Vardi et al., 2013b)).

It is evident that the total delay stretching of a five-neuron chain is superior to that of a two-neuron chain, as the stretching of each individual neuron is accumulative. The experimentally corroborated paradigm presented below is based on this key feature of the unavoidable accumulated stretching, enabling the implementation of different types of DLGs in the brain.

### 3. Experimentally examined DLGs

Neuronal logic-gates consist of a multilayer feedforward neural network, with a single output neuron. In this study we differentiate between two main classes of logic-gates, SLGs and DLGs. For illustration, a typical static neuronal AND-gate would consist of two input neurons and an output neuron which fires if and only if both input neurons are stimulated simultaneously. However, a dynamic AND-gate would change its functionality over time.

#### 3.1. Dynamic AND-Gate

The first experimentally examined feedforwad neuronal circuit is a dynamic AND-gate consisting of 5 neurons and 6 conditional stimulations, which split to weak/strong stimulations represented by dashed/full lines (**Figure 2A** (Vardi et al., 2013b)). A strong stimulation (above threshold) is characterized by a high amplitude and/or long duration, resulting in a reliable response. In contrary, a weak stimulation (sub threshold) is characterized by a lower amplitude and/or shorter duration, resulting in an evoked spike only in case of spatial or temporal summation, where the time-lag between two consecutive weak stimulations is short enough, as discussed in (Vardi et al., 2013b).

The delay of the three-neuron chain, $\tau_{AE}$, is defined as the time gap between stimulation to the input neuron and its corresponding stimulation to the output neuron (and similarly for other neuronal chains composing the DLG). Consequently, the time gap between two stimulations of the output neuron is $|\tau_{AE}-\tau_{BE}|$. Initially, $\tau_{AE}$ is shorter in comparison to the one-neuron chain, $\tau_{BE}$. This ratio reverses as repeated simultaneous stimulations are given to the input neurons, A and B, and the neuronal response latencies increase (**Figure 2B** (Vardi et al., 2013b)). For each input stimulation **Figure 2B** (upper panel, (Vardi et al., 2013b)) presents the time-lag between the two weak stimulations of neuron E, $|\tau_{AE}-\tau_{BE}|$, as well as whether a spike was evoked from neuron E. For a time-lag $|\tau_{AE}-\tau_{BE}|$ larger than ~0.5ms (varies among different neurons and stimulation parameters) the output neuron (E) does not respond, independent of the input stimulation, indicating a 'NULL' operating mode of the logic-gate. In the intermediate region, $|\tau_{AE}-\tau_{BE}|$ smaller than ~0.5 ms, the input/output interrelations typically follow that of an AND-gate. Hence, this neuronal gate exhibits NULL-AND-NULL dynamic logic transitions (**Table 1**, 1$^{st}$ row).

**Table 1 | Experimentally examined DLGs and their dynamic operations.** The first column lists the logic-gates. The second column details the truth table, the input/output relations. The third column presents the confirmed dynamic transitions among different logic operating modes, as a gate was repeatedly stimulated. The symbols "0/1" stand for a non-evoked/evoked spike, "NULL" indicates a non-evoked output spike independent of the inputs and IF($in_i$) indicates an output identical to the i$^{th}$ input. The order of IF($in_1$) and IF($in_2$) in the second row indicates the timing of their effects on the output unit. Reproduced upon permission by(Vardi et al., 2013b).

| Logic-gate | Truth table | | | Dynamic logic operation |
|---|---|---|---|---|
| | $in_1$ | $in_2$ | output | |
| AND | 0 | 0 | 0 | NULL → AND → NULL |
| | 0 | 1 | 0 | |
| | 1 | 0 | 0 | |
| | 1 | 1 | 1 | |
| OR | 0 | 0 | 0 | IF[$in_1$] + IF[$in_2$] → OR → IF[$in_2$] + IF[$in_1$] |
| | 0 | 1 | 1 | |
| | 1 | 0 | 1 | |
| | 1 | 1 | 1 | |
| NOT | 0 | | 1 | 1 → NOT → 1 |
| | 1 | | 0 | |
| XOR | 0 | 0 | 0 | OR → XOR → OR |
| | 0 | 1 | 1 | |
| | 1 | 0 | 1 | |
| | 1 | 1 | 0 | |

At the bottom of **Figure 2B** (Vardi et al., 2013b) different segments of the voltage recordings of neuron E are displayed, the colored (green, orange) lines are the stimulations arriving from the input chains ($\tau_{AE}$, $\tau_{BE}$, respectively). Initially, $\tau_{AE}$ is shorter than $\tau_{BE}$ (left recording) thus the "green" stimulation arrives at the output neuron before the "orange" one. This order is reversed later (right recording). The second and third recordings demonstrate the AND region; in the second recording two weak stimulations arriving at neuron E result in an evoked spike. In the case of response failure of one of the neurons comprising the left input chain (third recording), neuron E receives only one weak stimulation from neuron B and therefore does not fire, in agreement with the logic operation of an AND-gate (**Table 1**, 1$^{st}$ row).

The experimental results also indicate a slight asymmetry, where the first NULL-AND transition occurs at a shorter time-lag in comparison to the second AND-NULL transition (**Figure 2B** (Vardi et al., 2013b)). This asymmetry might be attributed to the stretching of the response latency of neuron E in between the two transitions.

### 3.2. Dynamic OR-Gate

The experimental setup of the dynamic OR-gate is similar to the AND-gate (**Figure 2A** (Vardi et al.,

2013b)), however all the stimulations are now strong (**Figure 3A** (Vardi et al., 2013b)) and are individually capable of reliably generating an evoked spike. The output neuron, F, generates two evoked spikes when the time-lag between the two incoming stimulations is large enough (compared to the refractory period), typically greater than 4 ms (**Figure 3B** (Vardi et al., 2013b)). To enhance the dynamic range of time-lags between two stimulations to neuron F, the gate now consists of six neurons in total and a four-neuron input chain (**Figure 3A** (Vardi et al., 2013b)). Consequently, the relative stretching of the two input neuronal chains, $|\tau_{AF}-\tau_{BF}|$ exceeds ~5 ms (**Figure 3C** and **Figure 3D** (Vardi et al., 2013b)).

The dynamic logic operating modes are exemplified for an entry from a region of typically two evoked spikes (when both input neurons are stimulated) into an OR mode, characterized by a single output spike in response to stimulation in $in_1$ *OR* $in_2$ (**Figure 3C** (Vardi et al., 2013b)), and for an exit from an OR mode (**Figure 3D** (Vardi et al., 2013b)). In the entry to the OR operating mode, the stimulation from neuron A (green) arrives prior to the stimulation from neuron B (orange), whereas in the exit, the "orange" stimulation arrives prior to the "green" one, and accordingly the order of the logic operations is presented in **Table 1**, $2^{nd}$ row. Note that *OR* represents one logic operation with one possible evoked spike, whereas the response of the DLG at the beginning/end is composed of 2 consecutive temporally independent logic operations. This can also be seen in the voltage recordings of neuron F, **Figure 3C** and **Figure 3D** (Vardi et al., 2013b).

### 3.3. Dynamic NOT-Gate

The implementation of the dynamic NOT-gate is similar to the previous ones (**Figure 2A** and **Figure 3A** (Vardi et al., 2013b)), however it contains an inhibitory stimulation from neuron D to E (**Figure 4A** (Vardi et al., 2013b)). It inhibits, for a limited time interval, the response of neuron E to an excitatory stimulation arriving from neuron B. Note that a typical NOT-gate consists of a single input (**Table 1**, $3^{rd}$ row), thus in our case the 'conventional' input is $in_1$. The inhibitory mechanism cannot be achieved by shaping the stimulation's amplitude or its sign. The use of a different cocktail of synaptic blockers, mainly suppressing the excitatory synapses (Appendix), enables the implementation of inhibitory stimulations, as discussed in (Vardi et al., 2013b). Since the effect of an inhibitory stimulation is measurable only in the presence of an excitatory stimulation, we apply an outer stimulation (indicated as $in_2$ in **Figure 4A** (Vardi et al., 2013b)) in the spirit of electronic circuits. This outer excitatory stimulation is applied each time a computation is requested, simultaneous with the stimulation of $in_1$.

For low stimulation rates, the stretching of neuronal response latencies is negligible; hence the logic operation of the gate was independently measured for each relative delay between excitation and inhibition of the output neuron E, $\tau_{BE}-\tau_{AD}$ (**Figure 4B** (Vardi et al., 2013b)) under low stimulation rate. When the inhibitory stimulation is given 5 ms or less prior to an excitatory stimulation - the inhibition is almost absolute. This effect deteriorates for larger time gaps, until it vanishes around 10 ms (**Figure 4B** (Vardi et al., 2013b)). For high stimulation rates, a dynamic behavior of the logic operation is demonstrated, where a relatively sharp transition is observed from a reliable relay of an arriving stimulation to an absolute blocker, a NOT-gate (**Figure 4C** (Vardi et al., 2013b)). In a reversed order, it is evident that an excitation sufficiently prior to inhibition is effective. However, it was experimentally difficult to locate this transition, since the spike detection is disrupted by the artifact of the inhibitory stimulation. Nevertheless, for an inhibition chain consisting of a larger number of neurons, consecutive 1-NOT-1 logic operating modes are anticipated in a single experiment (**Table 1**, 3$^{rd}$ row).

### 3.4. Dynamic XOR-Gate

The logic operation of a XOR-gate is identical to an OR-gate, except for the entry (1,1), two input stimulations, which do not generate an evoked spike (**Table 1**, 4$^{th}$ row). Its implementation is similar to the OR-gate setup with additional two inhibitory stimulations (Appendix), from the first input to a neuron belonging to the chain of the second input and vice versa (red connections in **Figure 4D** (Vardi et al., 2013b)). For low stimulation rates, the neuronal response latencies remain unaffected and the logical operation of the XOR-gate was tested independently for each relative delay between excitation and inhibition, $\tau_{BF}$–$\tau_{AC}$ (**Figure 4E** (Vardi et al., 2013b)). The delays $\tau_{AE}$ and $\tau_{BD}$ were selected such that the inhibition to neuron E is effective and consequently a transition from XOR to OR operating modes is exemplified (**Figure 4E** (Vardi et al., 2013b)). The confirmation of this dynamic logic operating transitions, however, requires much longer neuronal chains and is examined in section 4 using an analytical approach.

### 4. Theoretical analysis

Complex DLGs based on time-dependent neuronal response latencies usually require larger scale networks consisting of a greater amount of neurons. Their experimental implementations are associated with some difficulties, especially when delays, timing of stimulations and evoked spikes must be monitored on sub-millisecond timescales. Hence, the computational horizon of the new logic-gates requires a simplified theoretical framework which is based on the following two assumptions.

First, for each neuron comprising the gate, we assume a constant increase in the neuronal response latency per evoked spike, $\Delta$, independent of its current latency and identical for all neurons. This assumption approximately fits the second state of the latency increase (stimulation responses 100-650 in **Figure 1A** (Vardi et al., 2013b)). Under this assumption the latency of a neuron can be written as:

$$l(q) = l_0 + q\Delta \qquad (1)$$

where $l_0$ stands for the neuron's initial response latency, q is the number of evoked spikes and $\Delta$ is a constant which in our experiments is typically in the range of 2-7 μs. Similarly, the time delay of a chain is defined as the time-lag between the stimulation of the first neuron and the stimulation of the neuron at the end of the chain. Consequently the time delay for a chain consisting of n neurons is given by

$$\tau(q) = \tau_0 + nq\Delta \qquad (2)$$

where $\tau_0$ stands for the initial time delay of the chain. Similar to the experimental results, the increase in the delay of a chain is linear with the number of neurons in the chain, n.

The second assumption is that a strong excitatory stimulation generates an evoked spike with a probability of 1 (1:1 response), thus the number of evoked spikes of a neuron is equal to the number of its stimulations.

### 4.1. Dynamic AND-Gate

The AND-gate is examined below under this theoretical framework and results are compared to the experimental findings (section 3, **Figure 2A** (Vardi et al., 2013b)).

The delays of the green and orange chains (**Figure 2A** (Vardi et al., 2013b)) as a function of the stimulation number are presented in **Figure 5A** using equation (2) with $\Delta=5$ μs. The broadening of each line by 0.5 ms represents the maximal time delay between two stimulations of neuron E, $|\tau_{AE}-\tau_{BE}|$, which generates an evoked spike. Hence, the intersection between these two lines represents the region where neuron E fires. In agreement with the experimental results, the initial delay of the green chain (neurons A, C and D) is shorter than the delay of the orange chain (neuron B).

The similarities between the dynamical transition predicted by the theoretical model (**Figure 5B**) and the experimental results (**Figure 2B** (Vardi et al., 2013b)) are evident. Obviously, there are some minor differences; however the qualitative behavior is the same. This validation of the theoretical model supports its applicability for complex DLGs which are at the moment beyond experimental realization.

### 4.2. Generalized AND-Gate

Using the theoretical model presented above, several DLGs are examined. These DLGs implement complex transitions illustrating additional properties of their dynamics. To simplify the presentation we mainly concentrate on generalized AND-gates.

The first examined generalized AND-gate consists of three excitatory input chains consisting of 1/2/5 neurons (**Figure 6A**). A dashed arrow stands for a weak stimulation such that at least two weak stimulations at a time-lag less than 0.4 ms are required to generate an evoked spike in the output neuron. The initial time delays from the stimulations of the three input neurons to the stimulation of the output neuron are selected to be 30/27/25 ms for the chains consisting of 1/2/5 neurons, respectively. Note that in the limiting case of simultaneous stimulation to the three input neurons, this complex DLG is equivalent to the DLG consisting of only two input signals but with a more structured internal wiring, as exemplified in **Figure 6B**. Using equation (2) with $\Delta=0.004$ ms (4 μs) we show the time delays of the three input chains as a function of the number of given stimulations in **Figure 6C**. An intersection of two lines implies that the difference of the matching delays is less than 0.4 ms, thus resulting in a spike of the output neuron (black line in **Figure 6C**). In the intersection regions the gate acts as an AND gate for the two appropriate inputs (e.g., in the intersection of the "blue" and "orange" lines the output neuron fires if and only if $in_1$ *AND* $in_2$ are stimulated).

Increasing the input stimulation rate typically results in an enhanced stretching of the neuronal latency per spike (Vardi et al., 2012a). Results for $\Delta=0.006$ ms (6 μs) are presented in **Figure 6D**, where it is noticeable that the gate dynamics still consists of three entries to AND-regions. Moreover, the firing regions of **Figure 6C** and **Figure 6D** are the same under the rescaling of the stimulation axis by 0.004/0.006. Hence, we conclude that the dynamic transitions are robust to different stimulation frequencies. Nevertheless, it is clear that different initial delays to the three chains can reduce the three AND-reentries to two, one, or even remove the entire AND operation (e.g. the initial purple chain's delay is greater than the initial blue chain's delay which is greater than the initial orange chain's delay). Another important factor is the relative number of neurons comprising the neuronal chains. For illustration, in the case that the purple chain is reduced from five neurons (**Figure 6A**) to three, the three AND-regions merge into one region (**Figure 6E**).

In a more general scenario of k input chains to the output neuron where all input neurons are simultaneously stimulated, the maximal number of AND regions scales quadratically with k, since the number of intersections of k non-parallel lines is 0.5k(k-1). To exemplify a scenario where the number of transitions exceeds k, a gate with k=4 with 1/2/4/6 neuronal chains is examined (**Figure 7A**). Using equation (2) with $\Delta=0.006$ ms, where the maximal time-lag between two weak stimulations resulting in an evoked spike is 0.4 ms, one can spot six (0.5*4*3=6) transitions to an AND operating mode (**Figure 7A**).

To illustrate how the strength of the connections between neurons affects the gate's transitions, we examine an AND-gate of the same architecture as in **Figure 6A**, but the three input stimulations to the output neuron are weak and have the relative strengths of 0.3/0.75/0.5 for the orange/blue/purple connections, respectively (**Figure 7B**). To generate a spike at the output neuron, the sum of the stimulation strengths must exceed a threshold of 1. Note that the second transition to an AND-gate (**Figure 6A**) disappears (**Figure 7B**), since the sum of the strengths of the orange connection and purple connection is 0.8 and does not exceed the threshold (Markram and Tsodyks, 1996).

### 4.3. Dynamic XOR-Gate

The temporal activation of the XOR-gate was experimentally exemplified by a series of independent setups, where one of the inhibitory delays was gradually updated (**Figure 4E**). To illustrate the transitions of the dynamic XOR operation modes, three neurons are added to the excitatory purple input chain (**Figure 8**) in comparison to the experimental setup (**Figure 4D**). Initially we set the same delay for both inhibitions which are effective in a time window of [1,7] ms prior to the excitatory stimulation (i.e., if an inhibitory stimulation occurs at time T then the neuron will not respond to any stimulation in the time interval [T+1,T+7] ms). The region where the excitatory stimulation is inhibited is depicted by the light-red region bounded by dashed red lines (**Figure 8**). Consequently, $in_1$ is always inhibited by $in_2$, while $in_2$ is only temporarily inhibited by $in_1$, and a temporal XOR operation is observed.

### 4.4. Transition among multiple modes

In the following example we present a gate consisting of two inputs and an outer stimulation given for every computation (as in section 3.3, NOT-Gate), resulting in four different logic operating modes (**Figure 9**). The gate contains two inhibition chains (black and purple), with initial time delays of 30 and 42 ms, respectively. Both inhibitions are effective in a time window of [1, 7] ms prior to an excitatory stimulation (as in section 4.3). The initial blue and orange delays are 40 ms and 10 ms, respectively. For every computation of the logic-gate, the outer stimulation and the stimulations of the input neurons are given simultaneously. In the initial stage, the output neuron fires as a result of the outer stimulation independent of both inputs. The inhibition is ineffective, since the delays of the black and purple chains are too short (in comparison to the blue and orange delays). The black and purple delays increase with the neuronal response latencies, and the gate enters its second operating mode. The entire delay of the black chain grows relatively faster than the delay from the outer stimulation (blue) due to the number of neurons comprising each chain. Hence, when stimulated repeatedly, the delay of the black chain increases enough to inhibit the output spike which is caused by the outer stimulation, whereas the delay of the purple chain is still too short to affect the output. Consequently, the output spike caused by the excitatory outer stimulation is inhibited by $in_1=1$, resulting in a *NOT*($in_1$) functionality. In the third operation mode, the delays of the black and purple chains are both long enough to cause inhibition, therefore an output evoked spike will occur only in the case where both inputs are 0. In the fourth operation mode, the inhibition caused by the purple chain is still effective, whereas the inhibition caused by the black chain vanishes as a result of its enhanced stretching, resulting in a *NOT*(in$_2$) functionality. In the final operation mode, the delays of both inhibition chains are too large to inhibit the output spike caused by the outer stimulation, thus the logic-gate returns to its initial functionality where an output spike is generated independent of both inputs.

### 4.5. Varying inputs

So far, the limited case where simultaneous stimulations were given to all inputs of the gates was discussed. This scenario revealed many properties of the DLGs, however it is clear that more structured types of temporal input stimulations are expected to enrich the dynamic transitions. To exemplify this scenario we consider an AND-gate with two input chains consisting of 3 and 6 neurons (**Figure 10**). Applying a fixed stimulation rate to the two input neurons results solely in one AND-region (first AND region in **Figure 10**). A temporal reduction in the probability for a stimulation of the purple input chain results in a moderated latency increase, thus the delay of the blue chain becomes larger than the delay of the purple chain, and a second AND region emerges. When a fixed stimulation rate is applied again to the two input neurons, the delay of the purple chain overshoots the delay of the blue one, resulting in a reentry to a third AND region (**Figure 10**).

### 5. Multiple component networks and signal processing

We differentiate between two main computational capabilities of the DLGs. The first approach aims at reaching a specific operating mode of the dynamic gates using intentional repeated stimulations, which enables the desirable computations on occasional inputs. In the second approach, we are not interested in performing computations using specific logic operations but rather in using the dynamic properties of the gates. The purpose is to discover information regarding the input sequences. This approach is exemplified by a collaboration of a large number of dynamic components which together can implement a basic edge detector (**Figure 11**).

The input of an edge detector is a vector of size n and its task is to identify radical structural changes or discontinuities. For instance, if the vector's values represent a degree of brightness as a function of (one dimensional) position, the mission of an edge detector is to identify two consecutive points with significant changes in their brightness. The proposed edge detector, consisting of *n* input neurons, is sketched in **Figure 11**. Each two consecutive neurons serve as inputs to a dynamic AND-gate. Initially all delays are equal, thus simultaneous stimulations to all input neurons result in the firing of all output neurons. We assume that the number of stimulations of each input neuron is proportional to the brightness of the corresponding position in the input vector. To avoid extreme scenarios we assume that the inter-spike-intervals of each neuron do not vary much in time. Since the stretching of each delay is proportional to the number of input stimulations, a significant difference between two input chains of a dynamic AND-gate will be developed in case of a significant change between the brightness of two consecutive inputs. As a result their shared dynamic AND-gate will reach a NULL state. The examination of edges will be then achieved by a simultaneous stimulation to all input neurons. The sensitivity of the detection is determined by the duration of the stimulating period of the input neurons, where longer periods result in higher sensitivity. Since the stretching of the neuronal response latency is reversible, this edge detector can be reused after a short period without input stimulations.

## 6. Suitability of dynamic logic-gates to brain functionality

It is implausible to assume that brain functionality is as simple as a combination of standard SLGs, especially since it requires accurate predefined set of delays that are static and do not change over time. In this study we introduced a paradigm which is more suitable for brain functionality, DLGs. We will now discuss the feasibility and reliability of the DLGs in an environment more suitable for the functioning brain.

### 6.1. Short synaptic delays

Our experimental procedure, corroborated and extended by theoretical evidence, was examined under conditions of synaptic delays of a few tens of millisecond, which are typically beyond cortical synaptic delays. This constraint can be adapted to the time scales of synaptic delays and transient periods of the brain, several ms (Abeles, 1991). From a theoretical point of view, the functionality of the proposed feedforward logic-gates is a function of the relative difference between the stretching of the input chains, regardless of the absolute number of neurons constituting each chain. Therefore, all synaptic delays can be shortened to the order of a few milliseconds using long synfire chains (Abeles, 1991; 2004;Ikegaya et al., 2004;Izhikevich, 2006;Pastalkova et al., 2008;Long et al., 2010). For illustration, let us concentrate on **Figure 12A** consisting of relatively long delays, up to 34 ms. A similar modified dynamic gate consisting of long synfire chains, of 26/27/30 neurons (**Figure 12B**), resulting in 5-6 ms delays between consecutive neurons (including the neuronal response latency). Note that the relative difference between the amount of neuronal populations comprising the input synfire chains remain the same as in **Figure 12A**, i.e. 2-1=27-26 and 5-1=30-26. Therefore, both AND-gates have identical transition timings between NULL and AND logic operations.

### 6.2. Time scales of operation modes

The reported periods of operating logic modes consist of a few hundred stimulations, which exceed a few seconds under stimulation frequencies that are in the order of few dozens. These periods can be shortened by two orders of magnitude using the following two enhanced stretching effects: Long synfire chains increase the stretching linearly with the number of their relays and in addition, the neuronal response latencies increase significantly faster (by one order of magnitude) in the initial spiking activity (first state, **Figure 1A** (Vardi et al., 2013b)). Both of these biological ingredients are expected to significantly shorten mode's durations.

### 6.3. Population dynamics

The reliability of the DLGs is in question, since a finite probability of a neuronal response failure is expected. A mechanism to enhance signal-to-noise ratio can be achieved using population dynamics (Abeles, 1991; Buzsáki, 2010; Kanter et al., 2011; Kopelowitz et al., 2012). In a set of simulation studies composed of Hodgkin-Huxley neurons (Hodgkin and Huxley, 1952) at the population dynamics level we demonstrated that the time-dependent features of the new logic-gates remain valid. It is also expected that their functionality will become less sensitive to background fluctuations as the population representing each neuron increases (Vardi et al., 2012b). This feature is especially crucial to the realization of shorter synaptic delays, where the activity spontaneously terminates as a result of synaptic fatigue (Kawasaki et al., 2000; Ji et al., 2010) or neuronal refractory periods.

To examine the firing probability of a population stimulated by a sum of weak stimulations, we use the setup shown in **Figure 13A**. Populations A, B and C are comprised of 40 Hodgkin-Huxley neurons with parameters similar to those in (Kanter et al., 2011), where the synaptic reversal potential was set to be $E_{syn}$=0 and the maximal synaptic conductance for weak synaptic strengths, $g_{max}$, was set to 0.0662 mS/cm$^2$. Each neuron in population C was connected with probability of 0.1 to neurons in populations A and B, resulting in an average of 8 input stimulations for each neuron in population C. These diluted population-population stimulations, represented by the dashed arrows, are weak stimulations. Thus, to generate a spike in an output neuron, almost all stimulations from both populations A and B at a sufficiently small time-lag are required, as discussed in (Vardi et al., 2013b). The delays between neurons are taken from a Gaussian distribution with a standard deviation of 0.15 ms centered at $\tau_{AC}$ and $\tau_{BC}=\tau_{AC}+\gamma$, where $\gamma$ is the time lag between stimulations from populations A and B. The spiking probability of population C is measured as a function of the time-lag $\gamma$ (**Figure 13B**), indicating that for $\gamma$<1 ms more than half of the neurons comprising population C fire for a common drive to the input populations, A and B.

To demonstrate the dynamic AND-gate we construct a similar setup, containing a synfire chain from population B to population E (**Figure 13C**), where $g_{max}$=1.6 mS/cm$^2$ for strong synaptic strengths. The initial time delays between population are taken from a Gaussian distribution with a standard deviation of 0.15 ms. The neuronal response latency increase per evoked spike is taken to be $\Delta$=0.04 ms per spike (to reduce computation complexity). Simultaneous stimulations are given to all neurons in the input populations A and B. Initially, the difference $|\tau_{AE}-\tau_{BE}|$ is ~2 ms, therefore no output spikes are expected. As the delays between neuronal populations increase (as a result of the increase in the neuronal response latency of the population neurons) $|\tau_{AE}-\tau_{BE}|$ decreases, resulting in a population DLG, NULL-AND-NULL transitions (**Figure 13D**).

## 7. Conclusion

We proposed a new computational paradigm in which the brain consists of dynamic logic gates (DLGs) which are governed by time-dependent logic modes. The relevance of our work to the brain's functionalities has to be evaluated using many aspects including: (a) Do DLGs exist in the dynamics of a network of interconnected neurons? (b) Is the concept of DLGs robust to population dynamics and specifically to recurrent networks? (c) Is DLGs a mechanism which the brain could plausibly use to any extent and especially when it is critically rely on precise relative timing of neural activities? (d) Can one find a realistic learning mechanism, e.g. Hebb's rules, to implement DLGs?

The brain is composed of large neural networks, where neurons are interconnected via excitatory and inhibitory synapses as well as sub-threshold and above-threshold synapses. In the events of weak synapses, spatial and temporal summations of excitations are required to generate an evoked spike. Hence, the examined gate architectures have to be locally embedded in such large interconnected networks. The existence of weak synapses with high probability indicates that complex dynamic logic-gates, where several input chains exist, are also expected to be a common building block of such networks. We verified that the phenomenon of DLGs is robust to population dynamics and hence it is expected to be less sensitive to unexpected fluctuations in the response timings of a single neuron. However, there are many unavoidable effects of brain activity which are not assumed to carry any significant information, e.g. synaptic noise. Is the DLGs one of these unavoidable effects? The answer is not yet clear, however, we showed that the increase in the neuronal response latency to ongoing stimulations cannot be ignored, as it may double its value and therefore affect the time dependent connectivity of a recurrent network. As for the implication of such dynamic logic-gates to cognitive activities, we demonstrated some preliminary tasks such as edge detections, which obviously can be generalized to more complex tasks. Nevertheless, our work is a call for advanced in-vivo experiments and theoretical studies, which can pinpoint the existence and the importance of the suggested dynamic logic-gates in various functionalities of the brain. Moreover, the proposed mechanism of DLGs opens a manifold of theoretical questions regarding advanced paradigm for the brain activity including the search for efficient local learning rules for the DLGs.

It is evident that the variety of possible dynamic logic-gates is much larger than the abovementioned examples. For recurrent networks, the complexity is expected to be enhanced in comparison to feedforward networks. As opposed to feedforward networks with given simultaneous external stimulations, in recurrent networks the timings of the input stimulations are a function of the large scale activity of the entire network. One of the open theoretical questions is the number of realizable logic operations among $P^N$, where each one of the N gates has P operating modes.

On mathematical grounds, the key question is whether recurrent networks consisting of DLGs might go beyond the computation paradigm of the universal Turing machine (Turing, 1938; Maini et al., 2006; Dayan, 2009; Hodges, 2012). This challenge requires a careful mathematical definition and in particular, a definition of whether the stretching of the neuronal response latency has to be taken as continuous or discrete in comparison to the delays. Such networks represent a class of heterogeneous time-delayed networks composed of excitable units, where the delays are a function of the activity of the network itself. Practically, the question is whether a circuit composed of such new elements can be analyzed using the traditional systematic methods and tools developed for Boolean circuits. In the event that the presented dynamics is within traditional computational complexity, i.e. can be implemented using conventional computers, an interesting question is its advantages with respect to the implementation of the brain's functionalities.

## 8. Acknowledgments

We thank Moshe Abeles and Eytan Domany for fruitful discussions and comments on the manuscript, as well as the computational assistance by Mathias Mahn, Igor Reidler, Yair Sahar, Alexander Kalmanovitch and Haya Brama. The authors thank Hana Arnon for invaluable technical assistance. This research was supported by the Ministry of Science and Technology, Israel.

## 9. Appendix: Materials and methods

### 9.1. Culture preparation

Cortical neurons were obtained from newborn rats (Sprague -Dawley) within 48 h after birth using mechanical and enzymatic procedures (Marom and Shahaf, 2002). All procedures were in accordance with the National Institutes of Health Guide for the Care and Use of Laboratory Animals and Bar-Ilan University Guidelines for the Use and Care of Laboratory Animals in Research and were approved and supervised by the Institutional Animal Care and Use Committee. The cortical tissue was digested enzymatically with 0.05% trypsin solution in phosphate-buffered saline (Dulbecco's PBS) free of calcium and magnesium, supplemented with 20 mM glucose, at 37°C. Enzyme treatment was terminated with heat-inactivated horse serum, and cells were then mechanically dissociated. The neurons were plated directly onto substrate-integrated multi-electrode arrays (MEAs) and allowed to develop functionally and structurally mature networks over a time period of 2–3 weeks in vitro, prior to the experiments. Variability in the number of cultured days in this range had no effect on the observed results. The number of plated neurons in a typical network was in the order of 1,300,000, covering an area of about 380 mm$^2$. The preparations were bathed in minimal essential medium (MEM-Earle, Earle's Salt Base without L-Glutamine) supplemented with heat-inactivated horse serum (5%), glutamine (0.5 mM), glucose (20 mM), and gentamicin (10 g/ml), and maintained in an atmosphere of 37°C, 5% $CO_2$, and 95% air in an incubator as well as during the electrophysiological measurements. All experiments were conducted on cultured cortical neurons that were functionally isolated from their network by a pharmacological block of glutamatergic and GABAergic synapses. Experiments were conducted in the standard growth medium, supplemented with 10 μM CNQX (6-cyano-7-nitroquinoxaline-2,3-dione) and 80 μM APV (amino-5-phosphonovaleric acid). 5 μM Bicuculline was added only in experiments where no inhibitory stimulations were used (**Figures 1,2,3**). This cocktail of synaptic blockers made the spontaneous network activity sparse. At least one hour was allowed for stabilization of the effect.

### 9.2. Measurements and stimulations

An array of 60 Ti/Au/TiN extracellular electrodes, 30 μm in diameter, and spaced either 200 or 500 μm from each other (Multi-Channel Systems, Reutlingen, Germany) were used. The insulation layer (silicon nitride) was pre-treated with polyethyleneimine (0.01% in 0.1M Borate buffer solution). A commercial setup (MEA2100-2x60-headstage, MEA2100-interface board, MCS, Reutlingen, Germany) for recording and analyzing data from two 60-electrode MEAs was used, with integrated data acquisition from 120 MEA electrodes and 8 additional analog channels, integrated filter amplifier and 3-channel current or voltage stimulus generator (for each 60 electrode array). Mono-phasic square voltage pulses ([100,500] μs, [-900,-100] mV) were applied through extracellular electrodes. Each channel was sampled at a frequency of 50k sample/s. Action potentials were detected on-line by threshold crossing. For each of the recording channels a threshold for spike detection was defined separately, prior to the beginning of the experiment.

### 9.3. Cell selection

Each logic-gate's node was represented by a stimulation source (source electrode) and a target for the stimulation – the recording electrode (target electrode). These electrodes (source and target) were selected as the ones that evoked well-isolated, well-formed spikes and reliable response with high signal-to-noise ratio. This examination was done with stimulus intensity of -800 mV using 30 repetitions at a rate of 5Hz followed by 1200 repetitions at a rate of 10 Hz.

In experiments where inhibitory stimulations were used (NOT-gate, XOR-gate) Bicuculline was not added to the standard growth medium, hence inhibitory synapses were not blocked. The initial step to identify a pair of electrodes for an inhibitory stimulation was to pinpoint an excitatory node by its source and target electrodes (a stimulation of the source electrode, i, results in a detection of a well isolated spike in the target electrode, j). In the next step, stimulations were given to each one of the 60 extracellular electrodes (electrode k) a few ms prior to the stimulation of the source electrode, i, while the activity of the target electrode, j, was recorded. This procedure was repeated five times. This examination was performed under different time-lags between stimulations of electrode k (k=1 to 60) and the stimulation of the source electrode, i. In the case of an inhibitory stimulation (neuron k inhibits neuron j), a stimulation given to electrode k several ms prior to the stimulation of the source electrode (e.g. less than 7 ms, **Figure 14A** and **Figure 14B**) results in no neuronal response recorded by the target electrode, j. When the time-lag between the stimulations of electrode k and the source electrode is relatively long (e.g. 15 ms, **Figure 14C**), the inhibitory effect gradually disappears, and a spike will be detected in the target electrode.

### 9.4. Stimulation control

A node response was defined as a spike occurring within a typical time window of 2-10 ms following the electrical stimulation. The activity of all source and target electrodes was collected, and entailed stimuli were delivered in accordance to the connectivity of nodes in the logic-gate setup.

Gate connectivity, τ: Conditioned stimulations were enforced on the gate-neurons, embedded within a large-scale network of cortical cells in vitro, following the gate connectivity. Each gate delay is defined as the expected time between the spike and stimulation of two linked neurons; e.g. conditioned to a spike recorded in neuron i, a stimulation will be given to neuron j after $K_{ij}$ (ms). The time-lag between the stimulations of two linked neurons is defined as $\tau_{ij}$. Note that in the case of two neurons $\tau_{ij}=L_i+ K_{ij}$, where $L_i$ is the response latency of neuron i.

After an electrical stimulation is given to the output neuron of the gate (neuron E, F, E, G in **Figures 2A, 3A, 4A, 4D**, respectively (Vardi et al., 2013b)), the input neurons (A, B) are simultaneously stimulated again after a fixed delay. The longest path from the input neurons to the output neuron, together with the time-lag between a stimulation applied to the output neuron and the next stimulation of the input neurons, determine the stimulation frequency of all the neurons constituting the gate; e.g. initially in **Figures 2A** (Vardi et al., 2013b) the longest path from the input neurons to the output neuron is 80 ms, and for a 20 ms time-lag between the stimulation applied to the output neuron and the next stimulation of the input neurons the effective stimulation rate of the neuronal gate is ~10 Hz.

AND-gate: Strong stimulations, (-800 mV, 200 μs), which were given to all gate neurons excluding neuron E, result in a reliable neural response. Weak stimulations (-550 mV, 120 μs) were given to neuron E, such that an evoked spike is expected only if the time-lag between two consecutive weak stimulations is short enough. In cases where the time-lag between two consecutive stimulations was shorter than 100 μs (from the end of the first stimulation to the beginning of the consecutive one), a unified long stimulation (-550mV, 280 μs) was applied, to overcome technical limitations. All neurons were stimulated at a rate of 10 Hz.

OR-gate: Strong stimulations (-800 mV, 200 μs), resulting in a reliable neural response, were given to all gate neurons. All neurons were stimulated at a rate of 1 Hz (**Figure 3B** (Vardi et al., 2013b)) or 10 Hz (**Figure 3C** and **Figure 3D** (Vardi et al., 2013b)). Since for each input stimulation neuron F was stimulated twice, its effective stimulation rate in the case of two evoked spikes was 20 Hz (**Figure 3C** and **Figure 3D** (Vardi et al., 2013b)). This higher stimulation rate results in a deterioration of the neuronal response which screens the distinguishing effect of one or two evoked spikes. To prevent this discrepancy, neuron F was stimulated only every second round, such that its effective stimulation rate remains on the average 10 Hz.

NOT-gate: Strong stimulations (-800 mV, 200 μs) were given to all gate neurons, excluding neuron E and result in a reliable neural response. A weaker stimulation (-550 mV, 100 μs) was given to neuron E to enhance the inhibitory effect. All neurons were stimulated at a rate of 1 Hz (**Figure 4B** (Vardi et al., 2013b)) or 10 Hz (**Figure 4C** (Vardi et al., 2013b)).

XOR-gate: Strong stimulations (-800 mV, 200 μs) were given to all gate neurons besides neurons E and F and result in a reliable neural response. Weaker stimulations (-550 mV, 100 μs) were given to neurons E and F to enhance the inhibitory effect. All neurons were stimulated at a rate of 1 Hz. To overcome the low probability to find two inhibitory stimulations in a given culture, the same source and target electrodes were assigned to nodes E and F, and the same inhibitory electrode was assigned to nodes C and D. These neurons were stimulated in a time-lag of 50 ms, and since the stimulation rate was 1 Hz there were no conflicts in terms of timing (e.g. latency stretching, refractory period, etc.).

## 9.5. Data analysis

Analyses were performed in a Matlab environment (MathWorks, Natwick, MA, USA). Action potentials were detected by threshold crossing. In the context of this study, no significant difference was observed in the results under threshold crossing or voltage minima for spike detection.

Since only a detection of spike in a certain neuron leads to a conditional stimulation of its linked neuron, there was a need to handle missed stimulations as well as missed evoked spikes. This was handled differently according to the nature of the gate:

Neuronal response latency (**Figure 1** (Vardi et al., 2013b)): In **Figure 1A** (Vardi et al., 2013b) the time-lags between the neuron's evoked spikes and the electrical stimulations are presented. Unconditional stimulations were given at a rate of 10 Hz, indicating that a stimulation is given every 100 ms whether a spike was detected or not. Stimulation instances not resulting in evoked spikes are not shown in the graph. In **Figure 1B** (Vardi et al., 2013b) and **Figure 1C** (Vardi et al., 2013b), the time-lag between the evoked spikes of the input and output neurons are presented. Only instances resulting in an evoked spike of the output neuron are shown.

AND-gate (**Figure 2B** (Vardi et al., 2013b)): Only instances where two stimulations were applied to the output neuron are shown, since one (or zero) stimulation will never generate an evoke spike (see stimulation control, AND-gate).

OR-gate (**Figure 3C** and **Figure 3D** (Vardi et al., 2013b)): Only instances where one or two stimulations were applied to the output neuron are shown. In this case even a single stimulation can evoke a spike (see stimulation control, OR-gate), and marked as '-1'.

NOT-gate (**Figure 4C** (Vardi et al., 2013b)): Only instances where both excitatory and inhibitory stimulations were applied to the output neuron are shown. The probability of an evoked spike of the output neuron is calculated only when the two stimulation types are applied (see stimulation control, NOT-gate).

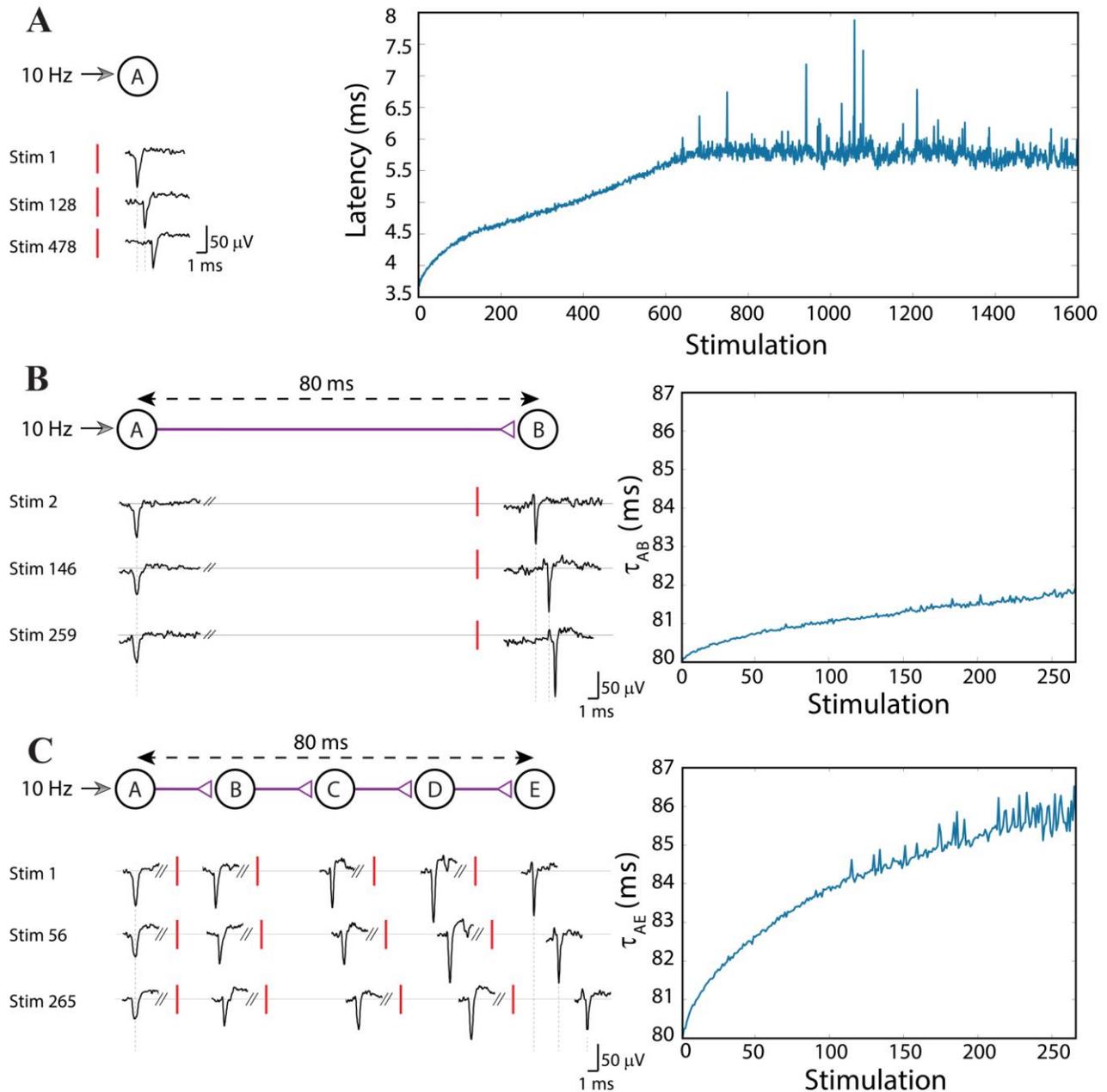

**Figure 1 (Color online) Stretching of the neuronal response latency to ongoing stimulations.** (**A**) An extracellular stimulation of a single neuron at a rate of 10 Hz. The relative time-gap between a stimulation (red bar) and its corresponding recorded evoked spike (voltage minima), the neuronal response latency, is exemplified for several stimulations (left). The graph (right) summarizes the response latencies over 1600 stimulations. (**B**) A two-neuron-chain where neuron A is stimulated at a rate of 10 Hz, and the initial delay between evoked spikes of neurons A and B is set to $\tau_{AB}=80$ ms. Several recorded spikes from neurons A and B are exemplified (left). The graph (right) summarizes the ~2 ms increase in $\tau_{AB}$ over ~270 stimulations. (**C**) Similar to **B** but with a five-neuron-chain, and a ~6 ms increase in $\tau_{AE}$ which accumulates the stretching of all four (B-E) neuronal response latencies. Reproduced upon permission from (Vardi et al., 2013b).

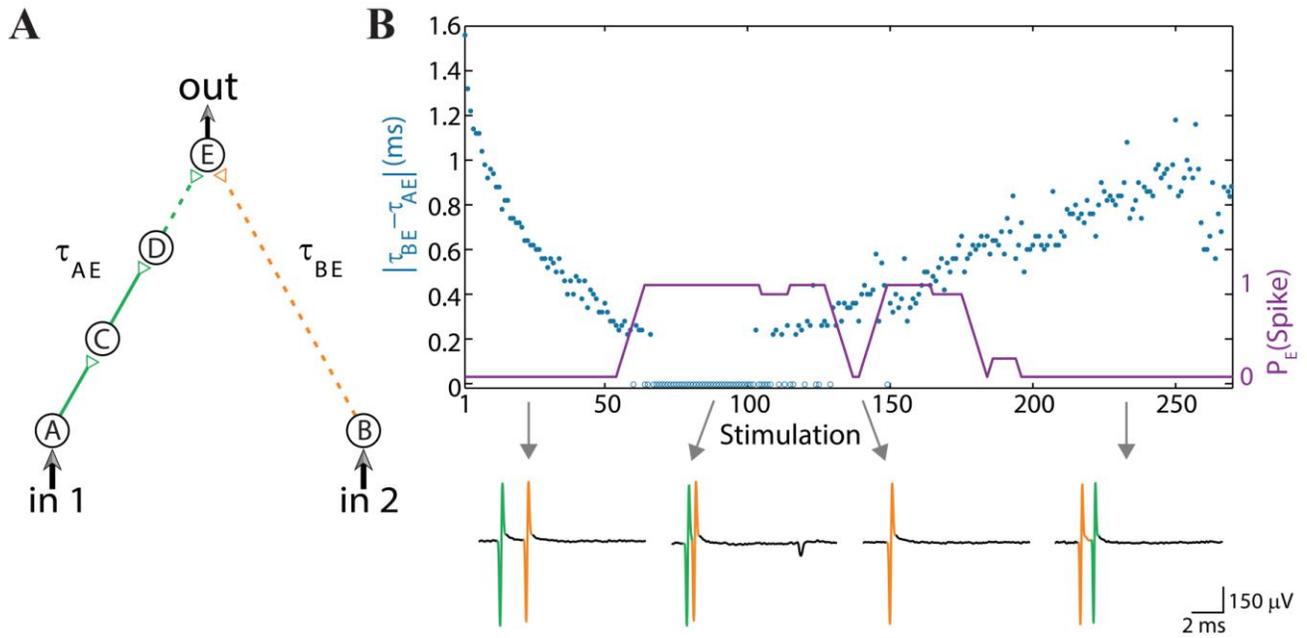

**Figure 2 (Color online) Dynamic AND gate.** (**A**) Schematic of an AND-gate consisting of five neurons and weak/strong stimulations (sub/above threshold) represented by dashed/full lines. (**B**) The delays are initially set to $\tau_{BE}$=80 ms and $\tau_{AE}\approx\tau_{BE}$-1.6 ms (in the presented experiment the initial delays between consecutive neurons in the left chain were selected to be equal, however, results are robust to arbitrary delays summing up to $\tau_{AE}$). Applying simultaneous stimulations at ~10 Hz to the input neurons, the two delays become the same and later reverse roles where $\tau_{AE}\approx\tau_{BE}$+1 ms, as presented by the blue circles as a function of the stimulation number. Unified longer stimulations were given for events where $|\tau_{AE}-\tau_{BE}|$<200 μs and are presented by zero time-lag open blue circles (Methods in Appendix). The probability of an evoked spike of neuron E over a sliding window of 10 stimulations is presented by the purple line. Different segments of the voltage recordings of neuron E are exemplified below, the arrows point from different scenarios to their matching recordings. Reproduced upon permission from (Vardi et al., 2013b).

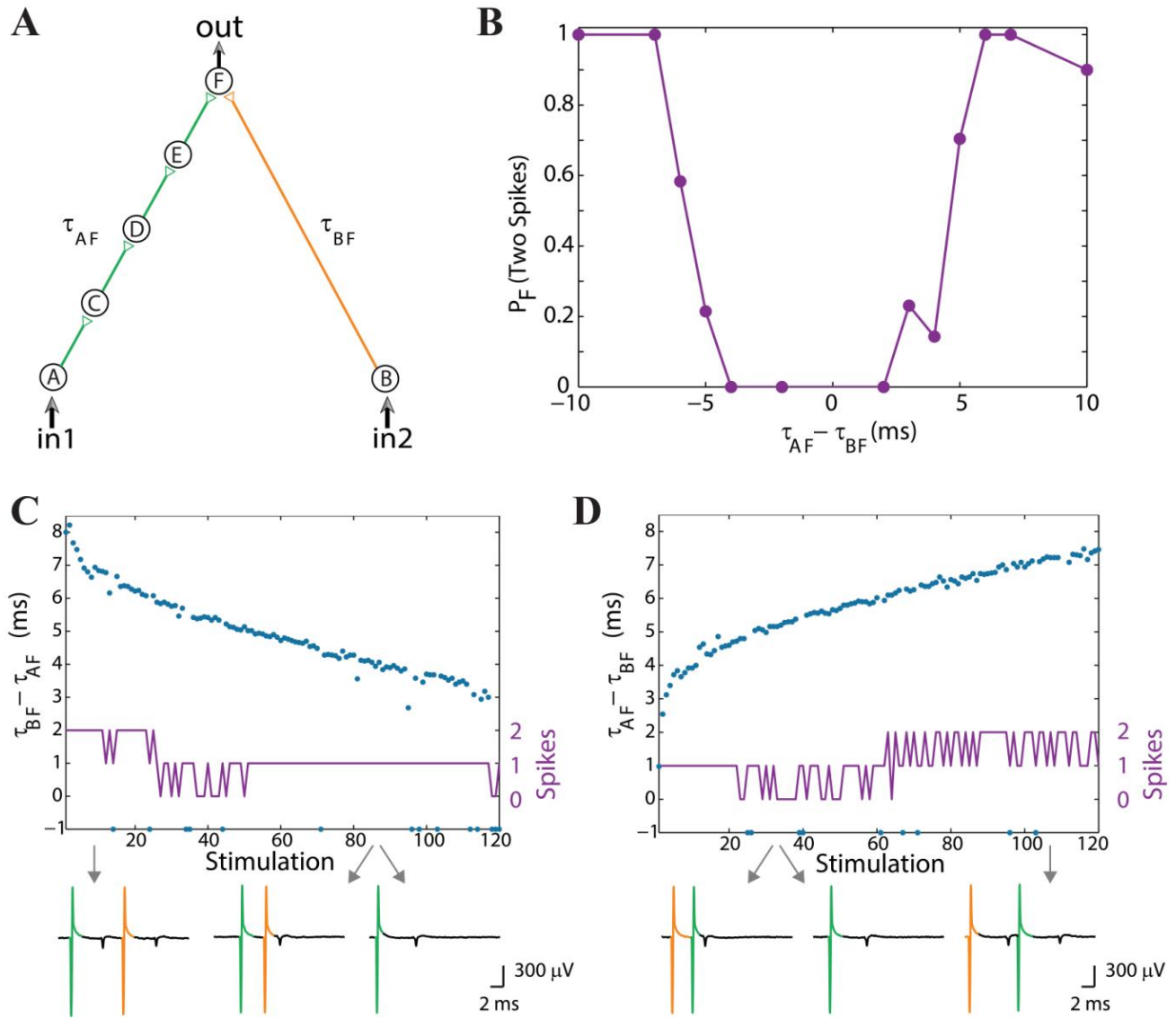

**Figure 3 (Color online) Dynamic OR gate.** (**A**) Schematic of an OR-gate consisting of a four-neuron input chain (green) and a one-neuron input chain (orange), where all stimulations are strong. (**B**) Independent experiments for a fixed time-lag $\tau_{AF}-\tau_{BF}$. The probability for neuron F to respond by two-spikes was averaged over several tens of input stimulations. (**C**) Input stimulations at a rate of 10 Hz resulting in dynamic changes of $\tau_{BF}-\tau_{AF}$ from 8 to 3 ms (blue dots). A dynamic transition from the region of typically two output spikes to an OR operating mode (similar to the entry in **B**) occurs after ~30 input stimulations. Missed evoked spikes resulting in only one stimulation to neuron F are marked as '-1'. (**D**) Similar to the entry in **B**, $\tau_{AF}-\tau_{BF}$ increases from ~2.5 to 7 ms (blue dots) and a dynamic exit from the OR region to the region of typically two evoked spikes occurs after ~60 input stimulations. Different segments of the voltage recording of neuron F are exemplified below, the arrows point from different scenarios to their matching recordings. Reproduced upon permission from (Vardi et al., 2013b).

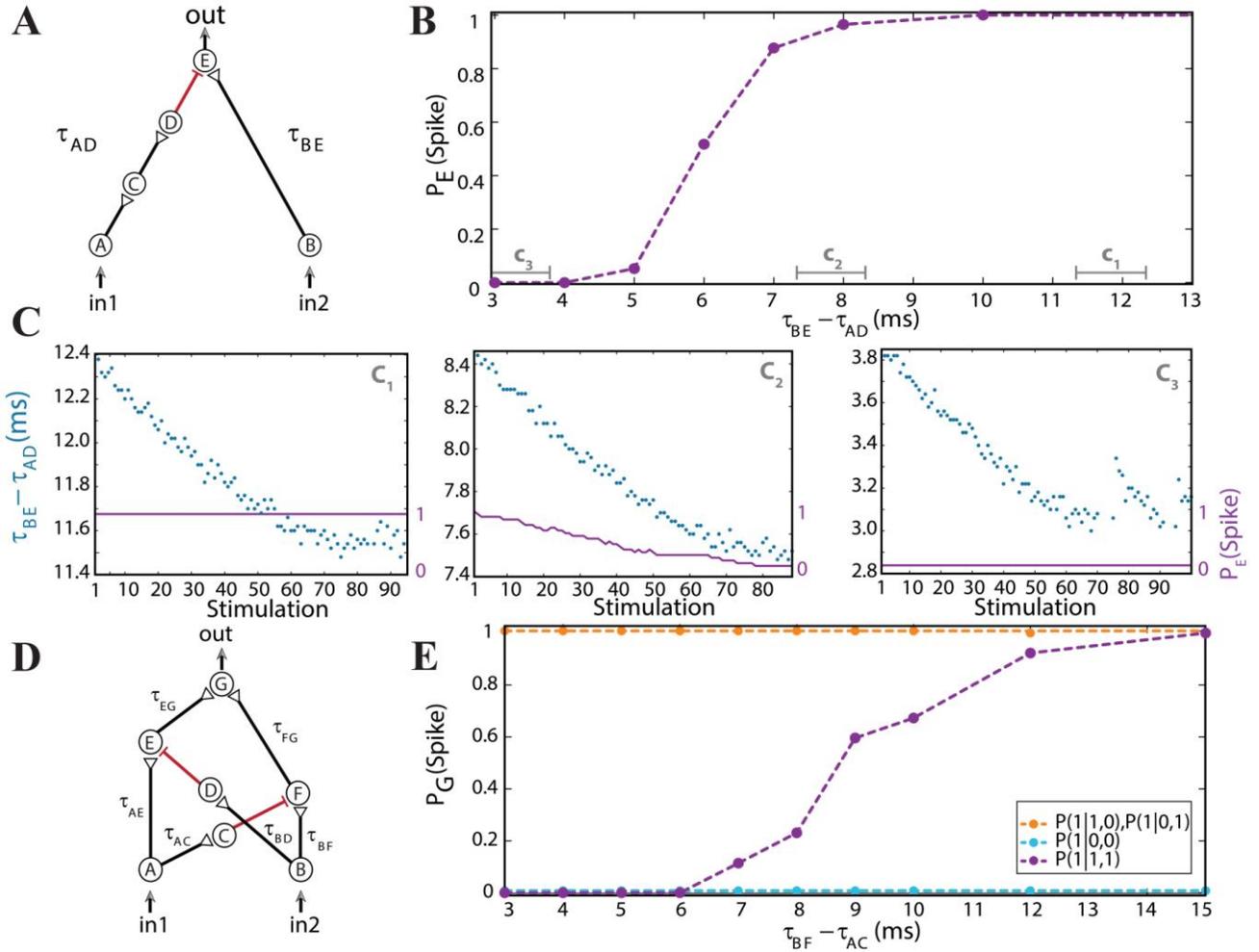

**Figure 4 (Color online) Dynamic NOT and XOR gates.** (**A**) Schematic of a NOT-gate consisting of five neurons, with one inhibition (red). A NOT-gate has one input (**Table 1**, 3$^{rd}$ row), where in$_2$ stands for an outer stimulation which is given for every computation. (**B**) Independent experiments for a fixed time-lag $\tau_{BE}$–$\tau_{AD}$ and $\tau_{BE}$=80 ms. The input neurons are simultaneously stimulated at 1 Hz. (**C**) Input stimulations at a rate of 10 Hz resulting in dynamic changes in $\tau_{BE}$–$\tau_{AD}$, averaged over a sliding window of 20 stimulations, as shown by time segments $c_1$, $c_2$ and $c_3$ in **B**. (**D**) Schematic of a XOR-gate containing two inhibitory stimulations (red). (**E**) Input neurons are simultaneously stimulated at 1 Hz. Independent experiments where $\tau_{BF}$–$\tau_{AC}$ is varied, a fixed time-lag $\tau_{AE}$–$\tau_{BD}$=3 ms was selected to inhibit the stimulation from neuron A, $\tau_{AE}$≈100 ms, $\tau_{BF}$≈50 ms and $\tau_{AG}$≈$\tau_{BG}$=150 ms were performed (circles connected with dashed guideline). The conditional probabilities of an evoked spike of the output neuron G are presented by the three colored dashed lines. Reproduced upon permission from (Vardi et al., 2013b).

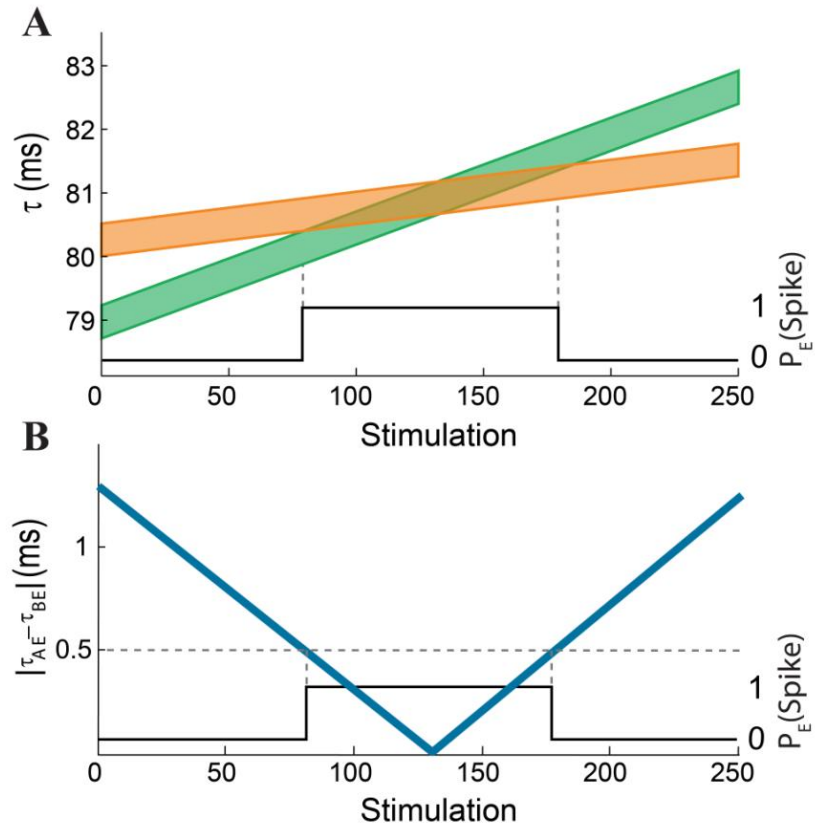

**Figure 5 (Color online) Theoretical analysis of the dynamic AND-gate.** (**A**) A graph of $\tau_{AE}$ (the lower border of the green line) and $\tau_{BE}$ (the lower border of the orange line) of the AND gate in **Figure 2B** as a function of the number of input stimulations. The width of the lines is 0.5 ms and the difference between the initial delays of the green and the orange chains is 1.3 ms. The black line indicates the firing probability of the output neuron. (**B**) The absolute difference between $\tau_{AE}$ and $\tau_{BE}$ as a function of the number of input stimulations (blue). The black line indicates the firing probability of the output neuron, similar to **Figure 2B**.

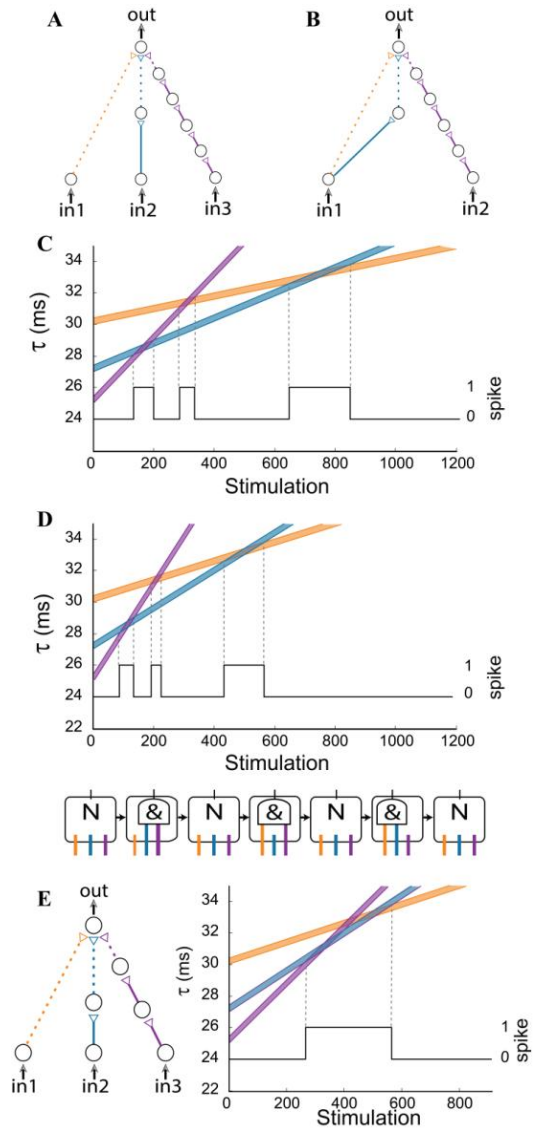

**Figure 6 (Color online) Generalized AND-gates exhibiting complex dynamic logic-gate transitions for simultaneous stimulations of all input neurons.** (**A**) Schematic of a generalized AND-gate consisting of three excitatory input chains. (**B**) Schematic of an AND-gate with two inputs which is equivalent to **A** for the case of simultaneous stimulations of the input neurons. (**C**) Time delays of the input chains as a function of stimulations, calculated using equation (2) for Δ=0.004 ms. The black line indicates the firing probability of the output neuron. (**D**) Same as **C** but the calculation is done for Δ=0.006 ms. Schematic of the equivalent time-dependent logic-gate is presented at the bottom, where a NULL ('N') operation stands for a non-evoked output spike independent of the input stimulations and '&' stands for an AND operation. (**E**) The same configuration and initial delays as in **D**, where the rightmost input chain (purple) is comprised now of 3 neurons (instead of 5). The three AND states merge into one region (bounded by two vertical dashed lines). The black line indicates the firing probability of the output neuron.

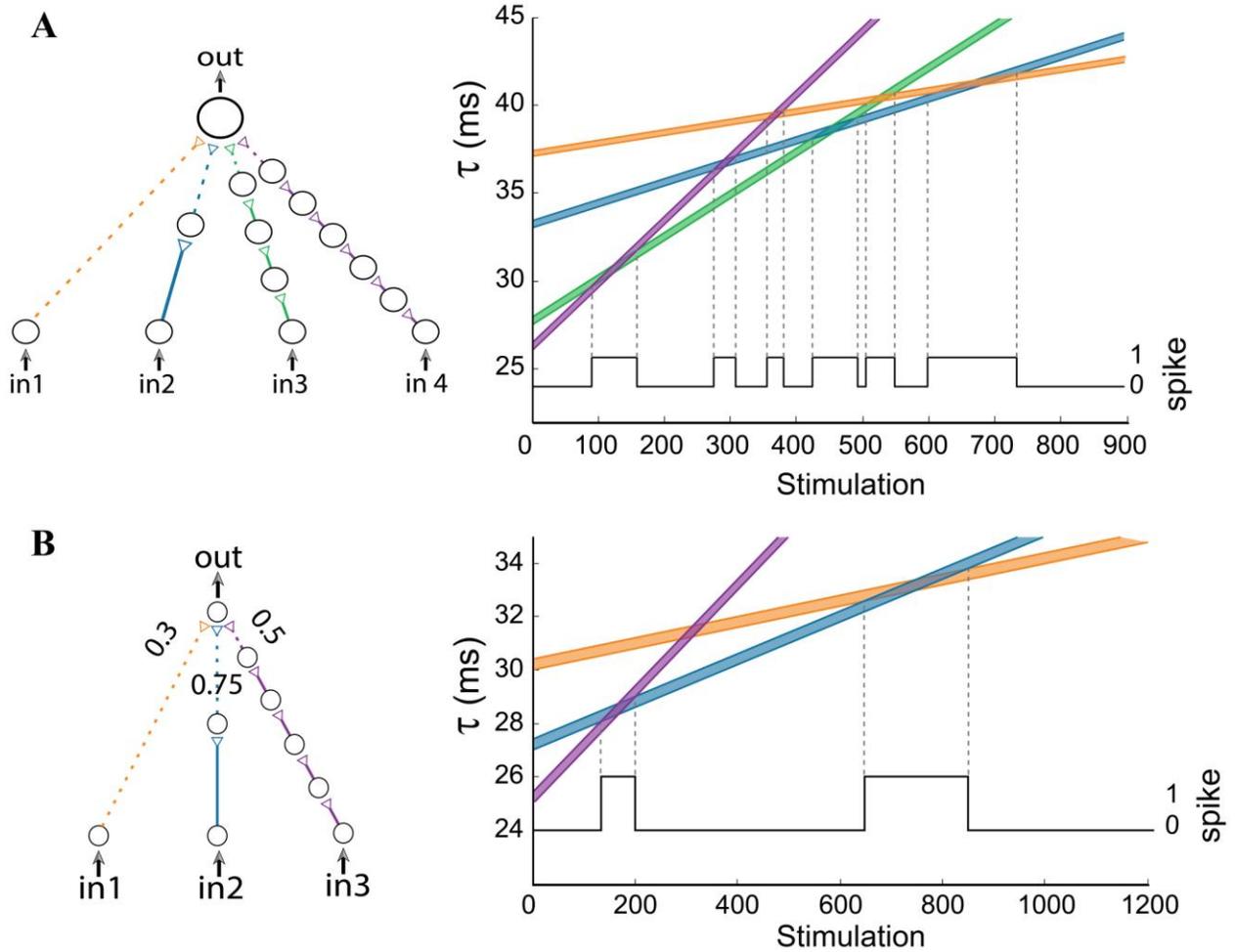

**Figure 7 (Color online) Advanced logic-gates.** (**A**) An AND-gate consisting of four inputs. The time delays of the input chains are presented as a function of the number of stimulations, calculated using equation (2) for Δ=0.006 ms. The black line indicates the firing probability of the output neuron. (**B**) An AND-gate of the same architecture as in **Figure 6A**, but the three weak stimulations have different strengths. The time delays of the input chains are presented as a function of the number of stimulations, calculated using equation (2) for Δ=0.004 ms. The black line indicates the firing probability of the output neuron.

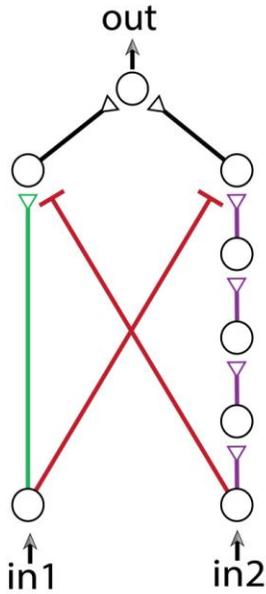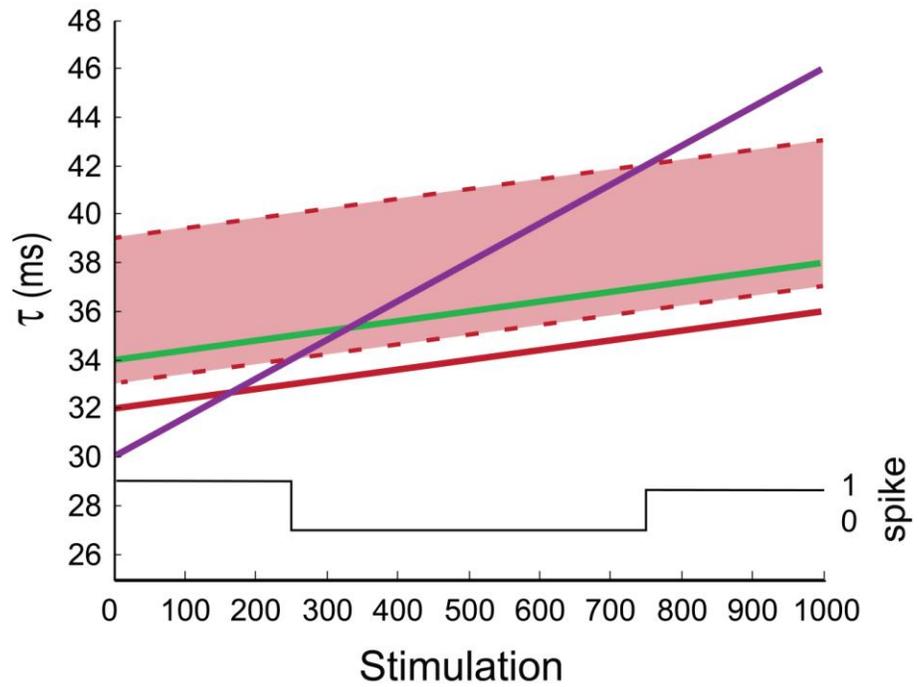

**Figure 8 (Color online) Dynamic XOR gate.** A dynamic XOR-gate with 2/5 neuronal excitatory input chains (green/purple), and two inhibitory stimulations (red) with identical initial delays of 32 ms. The inhibition is effective in a time window of [1,7] ms prior to the excitatory stimulation and is represented by the light-red region. The first input is always blocked (as the green line is always inside the light-red region). The black line indicates the firing probability of the output neuron. A temporal XOR operating mode is observed at the stimulation range of [250,750], where simultaneous stimulations (of $in_1$ and $in_2$) result in no evoked spikes of the output neuron.

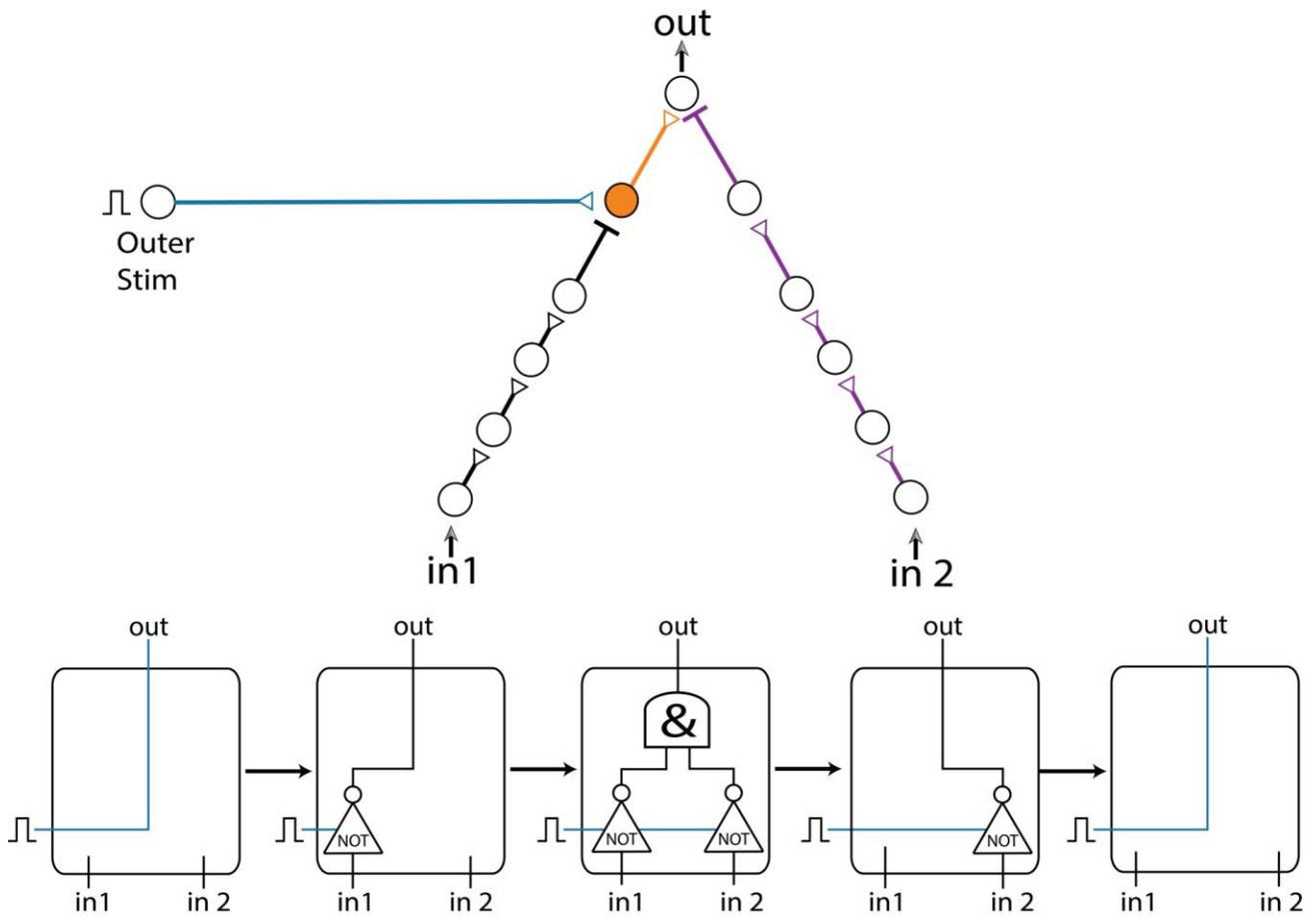

**Figure 9 (Color online) Multiple operation modes.** A gate consisting of two inputs, an outer stimulation and two inhibition chains (black and purple), exemplifying transitions among 5 different operation modes. The increase of the delays results in a transition between the logic operation modes illustrated by the flow chart at the bottom.

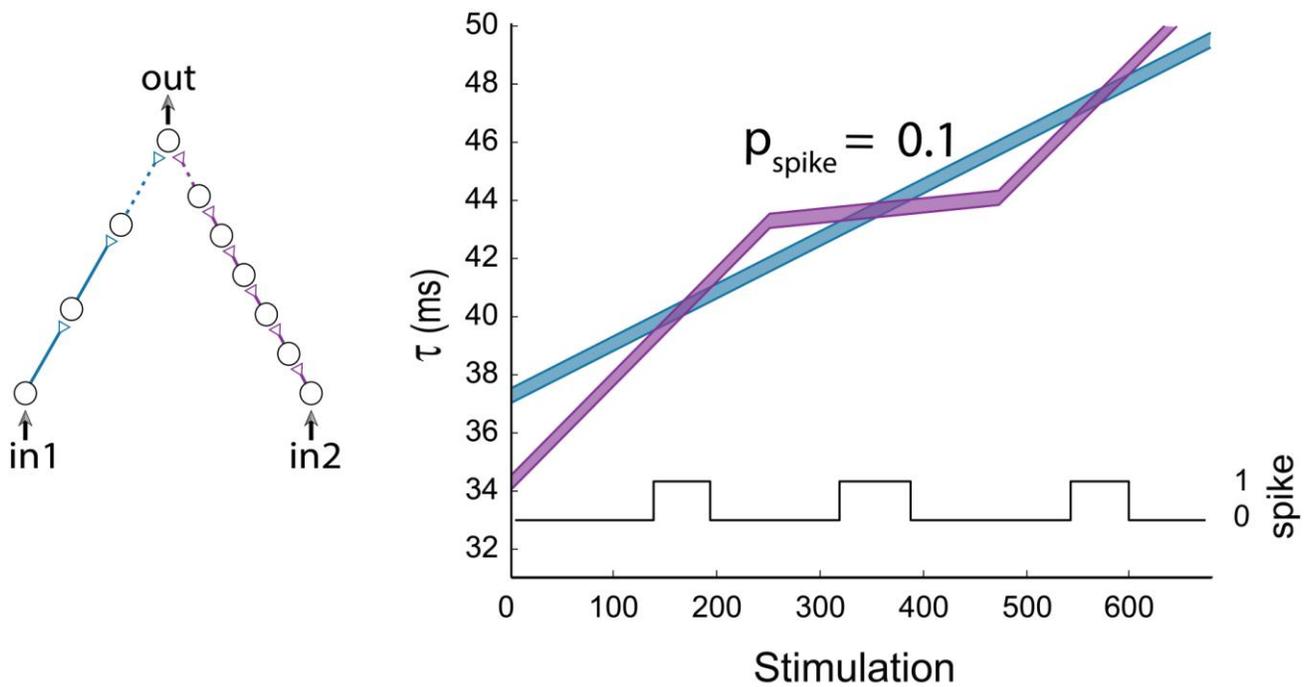

**Figure 10 (Color online) Non periodic input stimulations.** An AND-gate with the following input pattern: $in_1$ is stimulated at a fixed rate, while the stimulation of $in_2$ is relatively moderated in the stimulation period [250, 475] to probability 0.1 in comparison to $in_1$. The horizontal axis stands for the number of stimulation given to $in_1$. The black line indicates the firing probability of the output neuron per stimulation to $in_2$.

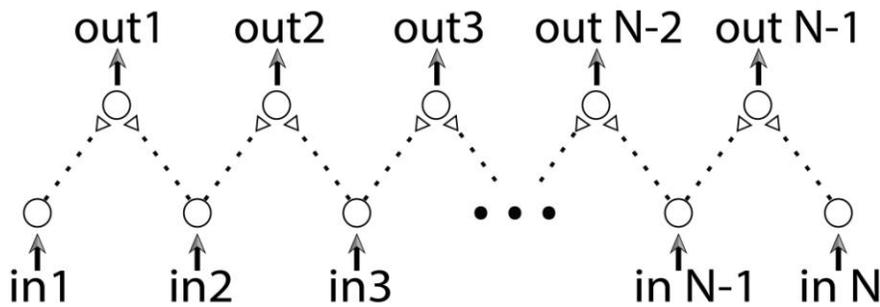

**Figure 11 (Color online) Edge detector.** An edge detector is built from a combination of dynamic AND-gates. The stimulation rate of each input neuron is proportional to the brightness of the corresponding position in the input vector.

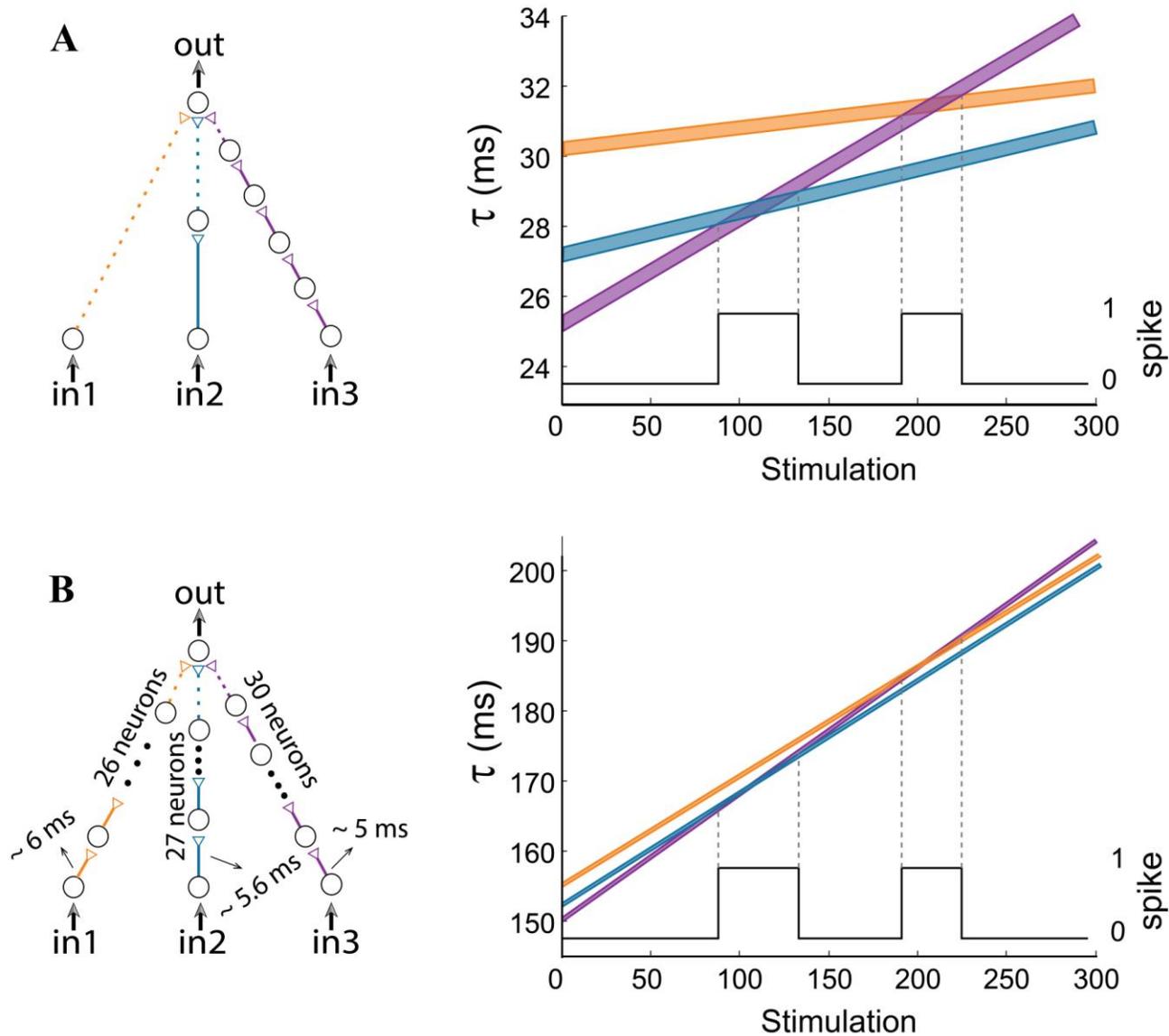

**Figure 12 (Color online) Short delays.** (**A**) The dynamic AND-gate and a portion of the graph presented in **Figure 6**, showing 2 transitions to AND regions. This gate consists of three input chains of 1/2/5 neurons each, and contains relatively long delays, up to 34 ms. (**B**) A similar AND-gate with long chains consisting of 26/27/30 neurons, resulting in short delays of 5-6 ms between consecutive neurons. For Δ=0.006 ms the delays of the input chains are presented in the right graph as a function of stimulation number, where the black line indicates the firing probability of the output neuron.

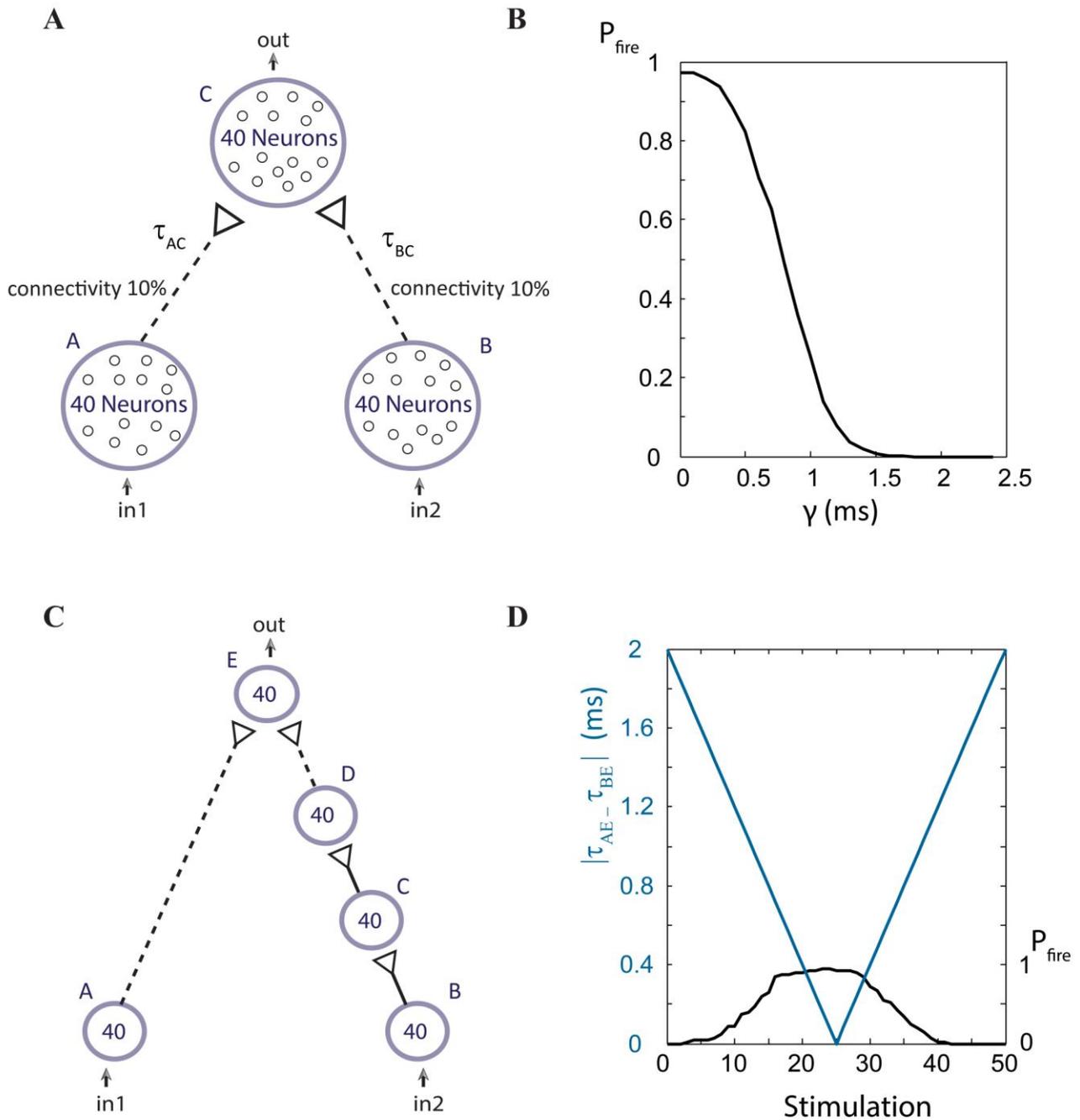

**Figure 13 (Color online) Population dynamics.** (**A**) Schematic of an AND-gate in population dynamics form. Population C receives week stimulations (represented by dashed arrows) from 0.1 of the neurons of each of the populations A and B. (**B**) For simultaneous stimulations of all neurons in populations A and B, the firing probability of the output population, C, is presented as a function of the time-lag between $\tau_{AC}$ and $\tau_{BC}$, $\gamma$. In the range where $\gamma$ is less than 1 ms an increased firing probability of population C is detected and the functionality of an AND-gate is maintained. (**C**) A dynamic AND-gate as in **Figure 2** in population dynamics form. (**D**) The input populations, A and B, are simultaneously stimulated, resulting in the decrease of the time-lag between stimulations of the output population $|\tau_{AE}-\tau_{BE}|$ (blue line) which increases again after ~25 stimulations. For short time-lags the output population fires at high probability (as shown in **B**) thus resulting in an AND mode functionality. For large time lags the probability is low and the gate is effectively NULL. Therefore, a dynamic NULL-AND-NULL transition is observed.

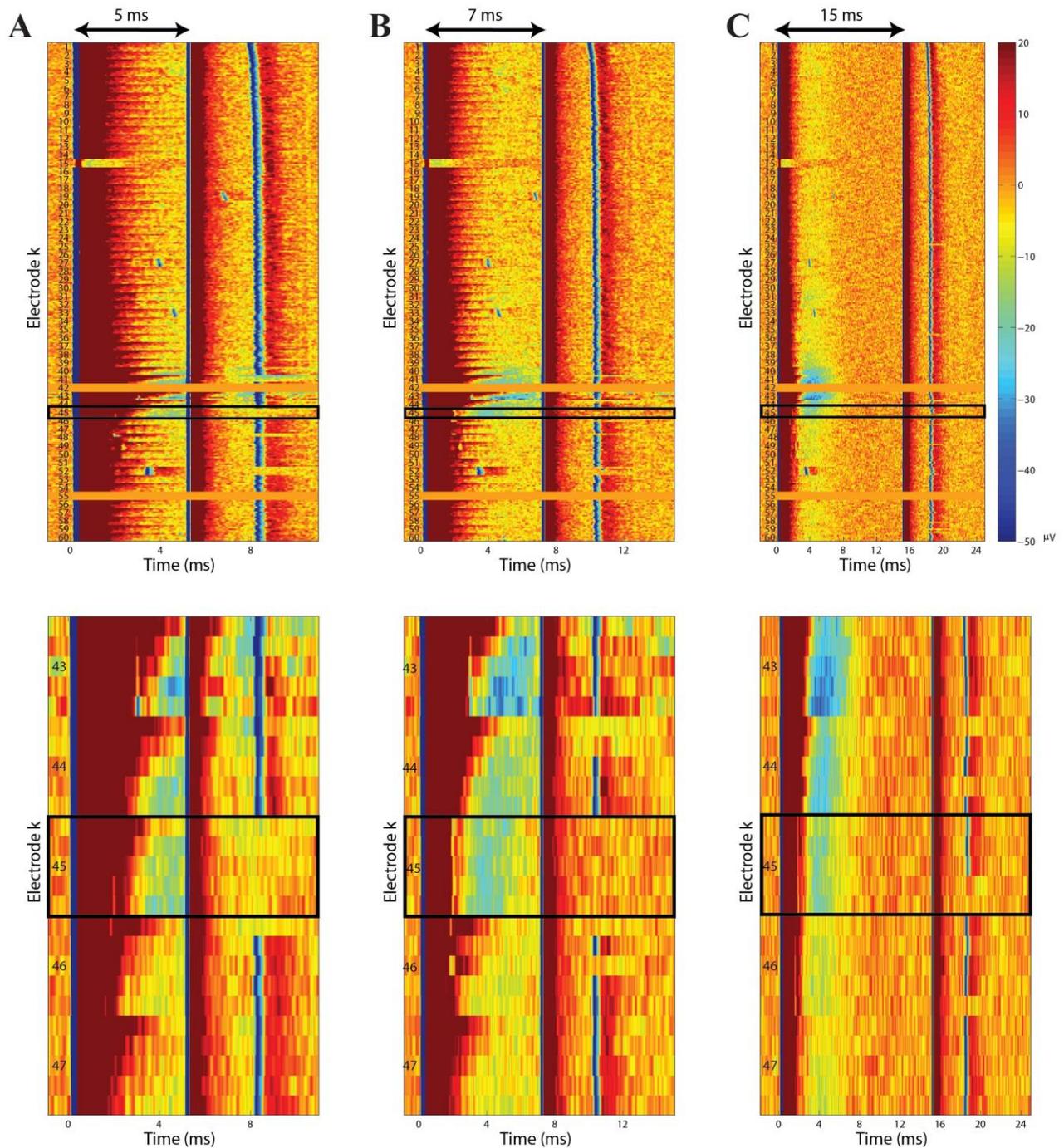

**Figure 14 (Color online) Inhibitory stimulations.** The voltage recorded from a neuron's target electrode (j=42) is presented in a color scale. Each row is independent from the others, and represents consecutive recordings. Row k represents the effect of an inhibitory stimulation (e.g. k=45, presented

enlarged at the lower panel) which precedes the stimulation of the neuron's source electrode (i=55). Stimulations of electrodes k=1 to 60 are given at time 0 (left dark blue column). Stimulations of the source electrode are given 5 ms (**A**), 7 ms (**B**) and 15 ms (**C**) after the stimulation of electrode k (middle dark blue column). The rightmost blue column represents the spikes recorded from the target electrode.

## 10. References

**Abeles, M. (1991).** *Corticonics: Neural circuits of the cerebral cortex.* Cambridge University Press.
**Aston-Jones, G., Segal, M., and Bloom, F.E. (1980).** Brain aminergic axons exhibit marked variability in conduction velocity. *Brain research* **195**, 215-222.
**Bakkum, D.J., Chao, Z. C. & Potter, S. M. (2008).** Long-term activity-dependent plasticity of action potential propagation delay and amplitude in cortical networks. *PLoS One* **3**, e2088.
**Ballo, A.W., and Bucher, D. (2009).** Complex intrinsic membrane properties and dopamine shape spiking activity in a motor axon. *J. Neuroscience* **29**, 5062-5074.
**Buzsáki, G. (2010).** Neural syntax: cell assemblies, synapsembles, and readers. *Neuron* **68**, 362-385.
**Chavesa, M., Albertb, R. & Sontaga, E. D. (2005).** Robustness and fragility of Boolean models for genetic regulatory networks. *J. Theo. Bio.* **235**, 431-449.
**Dayan, P. (2009).** A neurocomputational jeremiad. *Nat. Neuroscience* **12**, 1207-1207.
**De Col, R., Messlinger, K., and Carr, R.W. (2008).** Conduction velocity is regulated by sodium channel inactivation in unmyelinated axons innervating the rat cranial meninges. *J. physiology* **586**, 1089-1103.
**Eccles, J.C., Llinas, R., and Sasaki, K. (1966).** Excitatory Synaptic Action of Climbing Fibres on Purkinje Cells of Cerebellum. *J. Physiology* **182**, 268-296.
**Fuhrmann, G., Markram, H., and Tsodyks, M. (2002).** Spike frequency adaptation and neocortical rhythms. *J. Neurophysiology* **88**, 761-770.
**Gal, A., Eytan, D., Wallach, A., Sandler, M., Schiller, J., and Marom, S. (2010).** Dynamics of excitability over extended timescales in cultured cortical neurons. *J. Neuroscience* **30**, 16332-16342.
**Gerstner, W., Sprekeler, H., and Deco, G. (2012).** Theory and Simulation in Neuroscience. *Science* **338**, 60-65.
**Gilja, V., Nuyujukian, P., Chestek, C.A., Cunningham, J.P., Byron, M.Y., Fan, J.M., Churchland, M.M., Kaufman, M.T., Kao, J.C., and Ryu, S.I. (2012).** A high-performance neural prosthesis enabled by control algorithm design. *Nat. Neuroscience* **15**, 1752-1757.
**Grossman, Y., Parnas, I., and Spira, M.E. (1979).** Mechanisms involved in differential conduction of potentials at high frequency in a branching axon. *J Physiol.* **295**, 307-322.
**Hodges, A. (2012).** Beyond Turing's Machines. *Science* **336**, 163-164.
**Hodgkin, A.L., and Huxley, A.F. (1952).** A quantitative description of membrane current and its application to conduction and excitation in nerve. *J Physiol* **117**, 500-544.
**Hopfield, J.J. (1982).** Neural networks and physical systems with emergent collective computational abilities. *Proceedings of the national academy of sciences* **79**, 2554-2558.
**Hunt, L.T., Kolling, N., Soltani, A., Woolrich, M.W., Rushworth, M.F.S., and Behrens, T.E.J. (2012).** Mechanisms underlying cortical activity during value-guided choice. *Nat. Neuroscience* **15**, 470-476.
**Izhikevich, E.M. (2006).** Polychronization: computation with spikes. *Neural Comput.* **18**, 245–282
**Izhikevich, E.M., and Hoppensteadt, F.C. (2009).** Polychronous Wavefront Computations. *Int. J. Bifurcation and Chaos* **19**, 1733-1739.
**Ji, Y.Y., Lu, Y., Yang, F., Shen, W.H., Tang, T.T.T., Feng, L.Y., Duan, S.M., and Lu, B. (2010).** Acute and gradual increases in BDNF concentration elicit distinct signaling and functions in neurons. *Nat. Neuroscience* **13**, 302-309.

Kanter, I., Kopelowitz, E., Vardi, R., Zigzag, M., Kinzel, W., Abeles, M., and Cohen, D. (2011). Nonlocal mechanism for cluster synchronization in neural circuits. *EPL (Europhysics Letters)* **93**, 66001.

Kawasaki, F., Hazen, M., and Ordway, R.W. (2000). Fast synaptic fatigue in shibire mutants reveals a rapid requirement for dynamin in synaptic vesicle membrane trafficking. *Nat. Neuroscience* **3**, 859-860.

Kopelowitz, E., Abeles, M., Cohen, D., and Kanter, I. (2012). Sensitivity of global network dynamics to local parameters versus motif structure in a cortexlike neuronal model. *Physical Review E* **85**, 051902.

Krogh, A. (2008). What are artificial neural networks? *Nat. biotechnology* **26**, 195-197.

Litwin-Kumar, A., and Doiron, B. (2012). Slow dynamics and high variability in balanced cortical networks with clustered connections. *Nat. Neuroscience* **15**, 1498-1505.

Maini, P.K., Baker, R.E., and Chuong, C. (2006). The Turing model comes of molecular age. *Science* **314**, 1397.

Markram, H., and Tsodyks, M. (1996). Redistribution of synaptic efficacy between neocortical pyramidal neurons. *Nature* **382**, 807-810.

Marom, S., and Shahaf, G. (2002). Development, learning and memory in large random networks of cortical neurons: lessons beyond anatomy. *Quarterly reviews of biophysics* **35**, 63-87.

Mcculloch, W.S., and Pitts, W. (1943). A logical calculus of the ideas immanent in nervous activity. *The Bulletin of Mathematical Biophysics* **5**, 115-133.

Morin, F.O., Takamura, Y. & Tamiya, E. (2005). Investigating neuronal activity with planar microelectrode arrays: achievements and new perspectives. *J. bioscience and bioengineering* **100**, 131-143.

Nahin, P.J. (2012). *The Logician and the Engineer: How George Boole and Claude Shannon Created the Information Age.* Princeton Univ. Press.

Qian, L., Winfree, E. & Bruck, J. (2011). Neural network computation with DNA strand displacement cascades. *Nature* **475**, 368-372.

Rosenblatt, F. (1958). The perceptron: a probabilistic model for information storage and organization in the brain. *Psychological review* **65**, 386.

Scroggs, R.S. (2008). Evidence of a physiological role for use-dependent inactivation of nav1.8 sodium channels. *J Physiol* **586**, 923.

Shannon, C. (1938). A Symbolic Analysis of Relay and Switching Circuits. *Trans. AIEE* **57**, 713–723.

Soudry, D., and Meir, R. (2012). Conductance-based neuron models and the slow dynamics of excitability. *Front. Neurosci.* **6**, 4.

Spira, M.E., Yarom, Y. & Parnas, I. (1976). Modulation of spike frequency by regions of special axonal geometry and by synaptic inputs. *J Neurophysiol* **39**, 882–899.

Stoianov, I., and Zorzi, M. (2012). Emergence of a 'visual number sense' in hierarchical generative models. *Nat. neuroscience* **15**, 194-196.

Sutton, R.S., and Barto, A.G. (1998). *Reinforcement learning: An introduction.* Cambridge Univ Press.

Tal, D., Jacobson, E., Lyakhov, V. & Marom, S. (2001). Frequency tuning of inputoutput relation in a rat cortical neuron in-vitro. *Neurosci Lett* **300**, 21-24.

Thomson, A.M., and West, D.C. (1993). Fluctuations in Pyramid Excitatory Postsynaptic Potentials Modified by Presynaptic Firing Pattern and Postsynaptic Membrane-Potential Using Paired Intracellular-Recordings in Rat Neocortex. *Neuroscience* **54**, 329-346.

Turing, A.M. (1938). On computable numbers, with an application to the Entscheidungsproblem. A correction. *Proceedings of the London Mathematical Society* **2**, 544.

Van Pelt, J., Wolters, P.S., Corner, M.A., Rutten, W.L.C., and Ramakers, G.J.A. (2004). Long-term characterization of firing dynamics of spontaneous bursts in cultured neural networks. *Biomedical Engineering, IEEE Transactions on* **51**, 2051-2062.

Vardi, R., Goldental, A., Guberman, S., Kalmanovich, A., Marmari, H., and Kanter, I. (2013a). Sudden synchrony leaps accompanied by frequency multiplications in neuronal activity. *Frontiers in Neural Circuits* **7**, 176.

Vardi, R., Guberman, S., Goldental, A., and Kanter, I. (2013b). An experimental evidence-based

computational paradigm for new logic-gates in neuronal activity. *EPL (Europhysics Letters)* 103, 66001.

Vardi, R., Timor, R., Marom, S., Abeles, M., and Kanter, I. (2012a). Synchronization with mismatched synaptic delays: A unique role of elastic neuronal latency. *EPL (Europhysics Letters)* 100, 48003.

Vardi, R., Timor, R., Marom, S., Abeles, M., and Kanter, I. (2013c). Synchronization by elastic neuronal latencies. *Physical Review E* 87, 012724.

Vardi, R., Wallach, A., Kopelowitz, E., Abeles, M., Marom, S., and Kanter, I. (2012b). Synthetic reverberating activity patterns embedded in networks of cortical neurons. *EPL (Europhysics Letters)* 97, 66002.

Vogels, T.P., and Abbott, L.F. (2005). Signal propagation and logic gating in networks of integrate-and-fire neurons. *The Journal of neuroscience* 25, 10786-10795.

Von Neumann, J. (1956). Probabilistic logics and the synthesis of reliable organisms from unreliable components. *Automata studies* 34, 43-98.

Wagenaar, D., Pine, J., and Potter, S. (2006). An extremely rich repertoire of bursting patterns during the development of cortical cultures. *BMC neuroscience* 7, 11.